\documentclass[prd,aps,showpacs,epsf,floats,onecolumn]{revtex4}%
\usepackage{amssymb}
\usepackage{amsfonts}
\usepackage{amsmath}
\usepackage{graphicx}
\usepackage{fancyhdr}%
\providecommand{\U}[1]{\protect\rule{.1in}{.1in}}
\begin{document}
\title{\textbf{Comparing metrics for mixed quantum states: Sj\"{o}qvist and Bures}}
\author{\textbf{Paul M. Alsing}$^{1}$, \textbf{Carlo Cafaro}$^{2}$, \textbf{Orlando
Luongo}$^{3,4,5}$, \textbf{Cosmo Lupo}$^{6}$, \textbf{Stefano Mancini}$^{3,7}%
$, and \textbf{Hernando Quevedo}$^{5,8,9}$}
\affiliation{$^{1}$Air Force Research Laboratory, Information Directorate, 13441 Rome, New
York, USA}
\affiliation{$^{2}$SUNY Polytechnic Institute, 12203 Albany, New York, USA}
\affiliation{$^{3}$Universit\`{a} di Camerino, 62032 Camerino, Italy}
\affiliation{$^{4}$SUNY Polytechnic Institute, 13502 Utica, New York, USA}
\affiliation{$^{5}$Al-Farabi Kazakh National University, 050040 Almaty, Kazakhstan}
\affiliation{$^{6}$Politecnico di Bari, 70126 Bari, Italy}
\affiliation{$^{7}$INFN,\ Sezione di Perugia, 06123 Perugia, Italy}
\affiliation{$^{8}$Universidad Nacional Autonoma de Mexico, 04510 Mexico D. F., Mexico}
\affiliation{$^{9}$Universit\`{a} di Roma \textquotedblleft La Sapienza\textquotedblright,
00185 Roma, Italy}

\begin{abstract}
It is known that there are infinitely many distinguishability metrics for
mixed quantum states. This freedom, in turn, leads to metric-dependent
interpretations of physically meaningful geometric quantities such as
complexity and volume of quantum states.

In this paper, we first present an explicit and unabridged mathematical
discussion on the relation between the Sj\"{o}qvist metric and the Bures
metric for arbitrary nondegenerate mixed quantum states, using the notion of
decompositions of density operators by means of ensembles of pure quantum
states. Then, to enhance our comprehension of the difference between these two
metrics from a physics standpoint, we compare the formal expressions of these
two metrics for arbitrary thermal quantum states specifying quantum systems in
equilibrium with a reservoir at non-zero temperature. For illustrative
purposes, we show the difference between these two metrics in the case of a
simple physical system characterized by a spin-qubit in an arbitrarily
oriented uniform and stationary external magnetic field in thermal equilibrium
with a finite-temperature bath. Finally, we compare the Bures and Sj\"{o}qvist
metrics in terms of their monotonicity property.

\end{abstract}

\maketitle

\fancyhead[R]{\ifnum\value{page}<2\relax\else\thepage\fi}

\thispagestyle{fancy}

\section{Introduction}

The role played by geometric techniques in describing and, to a certain
extent, comprehending interesting classical and quantum physical phenomena of
relevance in Hamiltonian dynamics and statistical physics is becoming
increasingly important \cite{pettini07,cafarophysica22}. For instance, the
concepts of complexity \cite{felice18} and phase transition \cite{carollo20}
are two illustrative examples of physical phenomena being intensively
investigated with tools of information geometry \cite{amari}, i.e.
differential geometry combined with probability calculus. For example, the
singularities of a metric tensor of a manifold of coupling constants that
parametrize a quantum Hamiltonian can be shown to be linked to the quantum
phase transitions specifying the corresponding physical system
\cite{zanardi06,zanardi07,zanardi07B}. Moreover, the induced curvature of the
parameter manifold constructed from the metric tensor can also be viewed to
encode relevant information on peculiar characteristics of the system.
Specifically, the change in sign of the curvature, its discontinuities and,
finally, its possible divergences can be argued to be associated with
different (critical) regions of the parameter manifold where the statistical
properties of the physical system exhibit very distinctive behaviors
\cite{silva21,silva21B,mera22}.

In this paper we focus on the physics of quantum systems specified by mixed
quantum states because there exist infinitely many distinguishability
distances for mixed quantum states \cite{karol06}. This freedom in the choice
of the metric implies that these geometric investigations of physical
phenomena are still open to metric-dependent interpretations since a unifying
and complete conceptual understanding of these geometric tool (along with
their connections to experimental observations) has yet to be achieved. In
particular, given the non-uniqueness of such distinguishability distances,
understanding the physical relevance of considering either metric remains a
goal of great conceptual and practical interest \cite{silva21,silva21B,mera22}%
. Furthermore, for a chosen metric, comprehending the physical significance of
its corresponding curvature is essential and deserves further investigation
\cite{zanardi07,zanardi07B,pessoa21}.

An information geometric theoretical construct has recently been discussed
\cite{cafaroprd22} to describe and, to a certain extent, comprehend the
complex behavior of evolutions of quantum systems in pure and mixed states.
The comparative study was probabilistic in nature, i.e., it involved a
complexity measure \cite{cafaroPhD,cafaroCSF} based on a temporal averaging
procedure along with a long-time limit, and it was limited to examining
expected geodesic evolutions on the underlying manifolds. More specifically,
the authors studied the complexity of geodesic paths on the manifolds of
single-qubit pure and mixed quantum states equipped with the Fubini-Study
\cite{provost80,wootters81,braunstein94} and the Sj\"{o}qvist metrics
\cite{erik20}, respectively. They analytically showed that the evolution of
mixed quantum states in the Bloch ball is more\ complex than the evolution of
pure states on the Bloch sphere. They also verified that the ranking based on
their proposed measure of complexity, representing the asymptotic temporal
behavior of an averaged volume of the region explored on the manifold during
system evolutions, agreed with the geodesic length-based ranking. Finally,
targeting geodesic lengths and curvature properties in manifolds of mixed
quantum states, they observed a softening of the complexity on the Bures
manifold (i.e., a manifold of density operators equipped with the Bures metric
\cite{bures69,uhlman76,hubner92}) compared to the Sj\"{o}qvist manifold.

Motivated by the above-mentioned importance of choosing one metric over
another one in such geometric characterizations of physical aspects of quantum
systems and, in addition, intrigued by the different complexity behaviors
recorded with the Sj\"{o}qvist and Bures metrics in Ref. \cite{cafaroprd22},
we report in this paper a complete and straightforward analysis of the link
between the Sj\"{o}qvist metric and the Bures metric for arbitrary
nondegenerate mixed quantum states. Our presentation draws its original
motivation from the concise discussion presented by Sj\"{o}qvist himself in
Ref. \cite{erik20}, and it relies heavily on the concept of decompositions of
density operators by means of ensembles of pure quantum states
\cite{hughston93}. To physically deepen our understanding about the
discrepancy between these two metrics, we provide a comparison of the exact
expressions of these two metrics for arbitrary thermal quantum states
describing quantum systems in equilibrium with a bath at non-zero temperature.
Finally, we clarify the difference between these two metrics for a simple
physical system specified by a spin-qubit in an arbitrarily oriented uniform
and stationary external magnetic field vector in thermal equilibrium with a
finite-temperature environment.

The layout of the rest of the paper is as follows. In Section II, we revisit
the Sj\"{o}qvist metric construction for nondegenerate spectrally decomposed
mixed quantum states as originally presented in Ref. \cite{erik20}. In Section
III, inspired by the helpful remarks in Ref. \cite{erik20}, we make explicit
the emergence of the Bures metric from the Sj\"{o}qvist metric construction
extended to nondegenerate arbitrarily decomposed mixed quantum states. In
Sections II and III, we especially stress the role played by the concept of
geometric phase and the parallel transport condition for mixed states in
deriving the Sj\"{o}qvist and Bures metrics, respectively. In Section IV,
focusing on the physically relevant class of thermal quantum states and
following the works by Hubner in Ref. \cite{hubner92} and Zanardi and
collaborators in Ref. \cite{zanardi07B}, we cast the Sj\"{o}qvist and Bures
metrics in two forms suitable for an insightful geometric comparison between
the metrics. We end Section IV with a discussion of an illustrative example.
Specifically, we study the difference between the Sj\"{o}qvist and the Bures
metrics in the case of a physical system defined by a spin-qubit in an
arbitrarily oriented uniform and stationary external magnetic field in thermal
equilibrium with a finite-temperature environment. In Section V, we discuss
monotonicity aspects of the Sj\"{o}qvist metric. Our conclusive remarks along
with a summary of our main findings appear in Section VI. Finally, for ease of
reading, further technical details appear in Appendices A, B, and C.

\section{The Sj\"{o}qvist metric construction: Spectral decompositions}

In this section, we revisit the Sj\"{o}qvist metric construction for
nondegenerate spectrally decomposed mixed quantum states as originally
presented in Ref. \cite{erik20}. Before starting, we remark that the
Sj\"{o}qvist metric can be linked to observable quantities in suitably
prepared interferometric measurements. For this reason, it is sometimes termed
\textquotedblleft interferometric\textquotedblright\ metric
\cite{erik20,silva21}.

Let us consider two neighboring rank-$N$ nondegenerate density operators
$\rho\left(  t\right)  $ and $\rho\left(  t+dt\right)  $ specified by the
following ensembles of pure states,%
\begin{equation}
\rho\left(  t\right)  \overset{\text{def}}{=}\left\{  \sqrt{p_{k}\left(
t\right)  }e^{if_{k}\left(  t\right)  }\left\vert n_{k}(t)\right\rangle
\right\}  \text{, and }\rho\left(  t+dt\right)  \overset{\text{def}%
}{=}\left\{  \sqrt{p_{k}\left(  t+dt\right)  }e^{if_{k}\left(  t+dt\right)
}\left\vert n_{k}(t+dt)\right\rangle \right\}  \text{,} \label{cosmo1}%
\end{equation}
respectively, with $1\leq k\leq N$. Assume that $\left\langle n_{k}%
(t)\left\vert n_{k^{\prime}}(t)\right.  \right\rangle =\delta_{kk^{\prime}}$
and the phases $f_{k}\left(  t\right)  \in%
\mathbb{R}
$ for any $1\leq k\leq N$. Using Eq. (\ref{cosmo1}), $\rho\left(  t\right)  $
and $\rho\left(  t+dt\right)  $ can be recast in terms of their spectral
decompositions as%
\begin{equation}
\rho\left(  t\right)  =%
{\displaystyle\sum\limits_{k=1}^{N}}
p_{k}\left(  t\right)  \left\vert n_{k}(t)\right\rangle \left\langle
n_{k}(t)\right\vert \text{, and }\rho\left(  t+dt\right)  =%
{\displaystyle\sum\limits_{k=1}^{N}}
p_{k}\left(  t+dt\right)  \left\vert n_{k}(t+dt)\right\rangle \left\langle
n_{k}(t+dt)\right\vert \text{,} \label{cosmo2}%
\end{equation}
respectively. The Sj\"{o}qvist metric $d_{\mathrm{Sj\ddot{o}qvist}}^{2}\left(
t\text{, }t+dt\right)  $ between the two mixed quantum states $\rho\left(
t\right)  $ and $\rho\left(  t+dt\right)  $ in Eq. (\ref{cosmo1}) is formally
defined as \cite{erik20},%
\begin{equation}
d_{\mathrm{Sj\ddot{o}qvist}}^{2}\left(  t\text{, }t+dt\right)
=\underset{\left\{  f_{k}\left(  t\right)  \text{, }f_{k}\left(  t+dt\right)
\right\}  }{\min}%
{\displaystyle\sum\limits_{k=1}^{N}}
\left\Vert \sqrt{p_{k}\left(  t\right)  }e^{if_{k}\left(  t\right)
}\left\vert n_{k}(t)\right\rangle -\sqrt{p_{k}\left(  t+dt\right)  }%
e^{if_{k}\left(  t+dt\right)  }\left\vert n_{k}(t+dt)\right\rangle \right\Vert
^{2}\text{,} \label{cosmo3}%
\end{equation}
that is, after some algebra,%
\begin{equation}
d_{\mathrm{Sj\ddot{o}qvist}}^{2}\left(  t\text{, }t+dt\right)  =2-2%
{\displaystyle\sum\limits_{k=1}^{N}}
\sqrt{p_{k}\left(  t\right)  p_{k}\left(  t+dt\right)  }\left\vert
\left\langle n_{k}(t)\left\vert n_{k}(t+dt)\right.  \right\rangle \right\vert
\text{.} \label{cosmo4}%
\end{equation}
It is important to point out that in transitioning from Eq. (\ref{cosmo3}) to
Eq. (\ref{cosmo4}), the minimum is obtained by choosing phases $\left\{
f_{k}\left(  t\right)  \text{, }f_{k}\left(  t+dt\right)  \right\}  $ such
that%
\begin{equation}
\dot{f}_{k}\left(  t\right)  dt+\arg\left[  1+\left\langle n_{k}\left(
t\right)  \left\vert \dot{n}_{k}\left(  t\right)  \right.  \right\rangle
dt+O\left(  dt^{2}\right)  \right]  =0\text{.} \label{condo1}%
\end{equation}
Recall that an arbitrary complex number $z$ can be expressed as $z=\left\vert
z\right\vert e^{i\arg(z)}$. Then, noting that $e^{\left\langle n_{k}\left(
t\right)  \left\vert \dot{n}_{k}\left(  t\right)  \right.  \right\rangle
dt}=1+\left\langle n_{k}\left(  t\right)  \left\vert \dot{n}_{k}\left(
t\right)  \right.  \right\rangle dt+O\left(  dt^{2}\right)  $ is such that
$\arg\left[  e^{\left\langle n_{k}\left(  t\right)  \left\vert \dot{n}%
_{k}\left(  t\right)  \right.  \right\rangle dt}\right]  =-i\left\langle
n_{k}\left(  t\right)  \left\vert \dot{n}_{k}\left(  t\right)  \right.
\right\rangle dt$, Eq. (\ref{condo1}) can be recast to the first order in $dt$
as%
\begin{equation}
\dot{f}_{k}\left(  t\right)  -i\left\langle n_{k}\left(  t\right)  \left\vert
\dot{n}_{k}\left(  t\right)  \right.  \right\rangle =0\text{.} \label{condo2}%
\end{equation}
Eq. (\ref{condo2}) is the parallel transport condition $\left\langle \psi
_{k}\left(  t\right)  \left\vert \dot{\psi}_{k}\left(  t\right)  \right.
\right\rangle =0$ with $\left\vert \psi_{k}\left(  t\right)  \right\rangle
\overset{\text{def}}{=}e^{if_{k}\left(  t\right)  }\left\vert n_{k}\left(
t\right)  \right\rangle $ associated with individual pure state paths in the
given ensemble that specifies the mixed state $\rho\left(  t\right)  $
\cite{aharonov87}. For completeness, we recall here that a state $\rho\left(
t\right)  =U\left(  t\right)  \rho\left(  0\right)  U\left(  t\right)  $
evolving in a unitary fashion is parallel transported along an arbitrary path
when at each instant of time $t$ the state $\rho\left(  t\right)  $ is in
phase with the state $\rho\left(  t+dt\right)  =U\left(  t+dt\right)
U^{\dagger}\left(  t\right)  \rho\left(  t\right)  U(t)U^{\dagger}\left(
t+dt\right)  $ at an infinitesimal later time $t+dt$. Moreover, the parallel
transport conditions for pure (with $\rho\left(  t\right)  =\left\vert
\psi\left(  t\right)  \right\rangle \left\langle \psi\left(  t\right)
\right\vert $) and mixed states evolving in a unitary way are given by
$\left\langle \psi\left(  t\right)  \left\vert \dot{\psi}\left(  t\right)
\right.  \right\rangle =0$ and \textrm{tr}$\left[  \rho\left(  t\right)
\dot{U}(t)U^{\dagger}\left(  t\right)  \right]  =0$, respectively
\cite{erik00}. For a discussion on the parallel transport condition for mixed
quantum states evolving in a nonunitary manner, we refer to Ref.
\cite{tong04}. Interestingly, using clever algebraic manipulations and
expanding to the lowest nontrivial order in $dt$, $d_{\mathrm{Sj\ddot{o}%
qvist}}^{2}\left(  t\text{, }t+dt\right)  $ in Eq. (\ref{cosmo4}) can be
rewritten as%
\begin{equation}
d_{\mathrm{Sj\ddot{o}qvist}}^{2}\left(  t\text{, }t+dt\right)  =\frac{1}{4}%
{\displaystyle\sum\limits_{k=1}^{N}}
\frac{dp_{k}^{2}}{p_{k}}+\sum_{k=1}^{N}\left\langle \dot{n}_{k}\left\vert
\left(  \mathrm{I}-\left\vert n_{k}\right\rangle \left\langle n_{k}\right\vert
\right)  \right\vert \dot{n}_{k}\right\rangle dt^{2}\text{,} \label{cosmo4B}%
\end{equation}
with $\mathrm{I}$ in Eq. (\ref{cosmo4B}) denoting the identity operator. It is
worth observing that $ds_{k}^{2}\overset{\text{def}}{=}\left\langle \dot
{n}_{k}\left\vert \left(  \mathrm{I}-\left\vert n_{k}\right\rangle
\left\langle n_{k}\right\vert \right)  \right\vert \dot{n}_{k}\right\rangle
dt^{2}$ in Eq. (\ref{cosmo4B}) can be expressed as $ds_{k}^{2}=\left\langle
\nabla n_{k}\left\vert \nabla n_{k}\right.  \right\rangle $ where $\left\vert
\nabla n_{k}\right\rangle \overset{\text{def}}{=}\mathrm{P}_{\bot}^{\left(
k\right)  }\left\vert \dot{n}_{k}\right\rangle $ is the covariant derivative
of $\left\vert n_{k}\right\rangle $ and $\mathrm{P}_{\bot}^{\left(  k\right)
}\overset{\text{def}}{=}\mathrm{I}-\left\vert n_{k}\right\rangle \left\langle
n_{k}\right\vert $ is the projector onto states perpendicular to $\left\vert
n_{k}\right\rangle $. Furthermore\textbf{,} $\sum_{k}ds_{k}^{2}$ is the
nonclassical contribution in $d_{\mathrm{Sj\ddot{o}qvist}}^{2}\left(  t\text{,
}t+dt\right)  $ and represents a weighted average of pure state Fubini-Study
metrics along directions defined by state vectors $\left\{  \left\vert
n_{k}\right\rangle \right\}  _{1\leq k\leq N}$. This weighted average, in
turn, can be regarded as a generalized version of the Provost-Vallee coherent
sum procedure utilized to define a Riemannian metric on manifolds of pure
quantum states in Ref. \cite{provost80}\textbf{.} The derivation of Eq.
(\ref{cosmo4}) ends our revisitation of the original Sj\"{o}qvist metric
construction for nondegenerate mixed quantum states. It is important to
emphasize that $d_{\mathrm{Sj\ddot{o}qvist}}^{2}\left(  t\text{, }t+dt\right)
$ in Eq. (\ref{cosmo4}) was obtained by using the spectral decompositions of
the two neighboring mixed states $\rho\left(  t\right)  $ and $\rho\left(
t+dt\right)  $. Therefore, the metric was calculated for a special
decomposition of neighboring density operators expressed in terms of ensembles
of pure states.

\section{The Sj\"{o}qvist metric construction: Arbitrary decompositions}

In this section, we make explicit the emergence of the Bures metric from the
Sj\"{o}qvist metric construction (presented in Section II) extended to
nondegenerate arbitrarily decomposed mixed quantum states. In particular, we
emphasize the role played by the concept of geometric phase and the parallel
transport condition for mixed states in this derivation of the Bures
metrics.\textbf{ }Our discussion is an extended version of the abridged
presentation in Ref. \cite{erik20}.

\subsection{From spectral to arbitrary decompositions}

It is well-known in quantum information and computation that a given density
matrix can be expressed in terms of different ensembles of quantum states. In
particular, the eigenvalues and eigenvectors of a density matrix just denote
one of many possible ensembles that may generate a fixed density matrix. This
flexibility leads to the so-called theorem on the unitary freedom in the
ensembles for density matrices \cite{nielsen00}. This theorem implies
that\textbf{ }$\rho=\sum_{i}p_{i}\left\vert \psi_{i}\right\rangle \left\langle
\psi_{i}\right\vert =\sum_{j}q_{j}\left\vert \varphi_{j}\right\rangle
\left\langle \varphi_{j}\right\vert $\textbf{ }for normalized\textbf{
}states\textbf{ }$\left\{  \left\vert \psi_{i}\right\rangle \right\}
$\textbf{ }and\textbf{ }$\left\{  \left\vert \varphi_{j}\right\rangle
\right\}  $\textbf{ }and probability distributions\textbf{ }$\left\{
p_{i}\right\}  $ and\textbf{ }$\left\{  q_{j}\right\}  $\textbf{ }if and only
if\textbf{ }$\sqrt{p_{i}}\left\vert \psi_{i}\right\rangle =\sum_{j}%
u_{ij}\left\vert \varphi_{j}\right\rangle $\textbf{ }for some unitary
matrix\textbf{ }$u_{ij}$\textbf{,} and we may fill the smaller ensemble with
zero-probability entries in order to get same-size ensembles. In what follows,
we shall see the effect on metrics for mixed quantum states produced by this
unitary freedom in the ensembles for density matrices.

Let us consider arbitrary decompositions of two rank-$N$ neighboring density
operators $\rho\left(  t\right)  $ and $\rho\left(  t+dt\right)  $ in terms of
statistical ensembles of pure states. Let us start by defining the following
set $\left\{  \left\vert s_{k}\left(  t\right)  \right\rangle \right\}
_{1\leq k\leq N}$ of quantum states%

\begin{equation}
\left\vert s_{k}\left(  t\right)  \right\rangle \overset{\text{def}}{=}%
\sqrt{p_{k}\left(  t\right)  }\left\vert n_{k}\left(  t\right)  \right\rangle
\text{,} \label{cosmo5}%
\end{equation}
with $\left\langle s_{k}\left(  t\right)  \left\vert s_{k}\left(  t\right)
\right.  \right\rangle =p_{k}\left(  t\right)  $ for any $1\leq k\leq N$.
Then, given $\rho\left(  t\right)  \overset{\text{def}}{=}\left\{
e^{if_{k}\left(  t\right)  }\left\vert s_{k}\left(  t\right)  \right\rangle
\right\}  $, the spectral decomposition of $\rho\left(  t\right)  $ is
\begin{equation}
\rho\left(  t\right)  =%
{\displaystyle\sum\limits_{k=1}^{N}}
\left\vert s_{k}(t)\right\rangle \left\langle s_{k}(t)\right\vert \text{.}
\label{cosmo6}%
\end{equation}
Consider a unitary matrix $V\left(  t\right)  $ satisfying the unitary
condition $V^{\dagger}\left(  t\right)  V\left(  t\right)  =V\left(  t\right)
V^{\dagger}\left(  t\right)  =I$, with $I$ being the $N\times N$ identity
matrix. In terms of complex matrix coefficients $\left\{  V_{hk}\left(
t\right)  \right\}  _{1\leq h\text{, }k\leq N}$, the unitary condition can be
expressed as%
\begin{equation}%
{\displaystyle\sum\limits_{h=1}^{N}}
V_{hk}\left(  t\right)  V_{hl}^{\ast}\left(  t\right)  =\delta_{kl}\text{.}
\label{cosmo7}%
\end{equation}
Using the set $\left\{  \left\vert s_{k}\left(  t\right)  \right\rangle
\right\}  _{1\leq k\leq N}$ in\ Eq. (\ref{cosmo5}) and the unitary matrix
$V\left(  t\right)  $, we define a new set of normalized state vectors
$\left\{  \left\vert u_{h}\left(  t\right)  \right\rangle \right\}  _{1\leq
h\leq N}$ as%
\begin{equation}
\left\vert u_{h}\left(  t\right)  \right\rangle \overset{\text{def}}{=}%
{\displaystyle\sum\limits_{k=1}^{N}}
V_{hk}\left(  t\right)  \left\vert s_{k}\left(  t\right)  \right\rangle
\text{.} \label{cosmo8}%
\end{equation}
Given the set $\left\{  \left\vert u_{h}\left(  t\right)  \right\rangle
\right\}  _{1\leq h\leq N}$ with $\left\vert u_{h}\left(  t\right)
\right\rangle $ in Eq. (\ref{cosmo8}), we observe that we have constructed a
set of unitarily equivalent representations of the mixed state $\rho\left(
t\right)  $. Indeed, we have%
\begin{align}
\sum_{h=1}^{N}\left\vert u_{h}\left(  t\right)  \right\rangle \left\langle
u_{h}\left(  t\right)  \right\vert  &  =\sum_{h,k,l=1}^{N}V_{hk}\left(
t\right)  V_{hl}^{\ast}\left(  t\right)  \left\vert s_{k}\left(  t\right)
\right\rangle \left\langle s_{l}\left(  t\right)  \right\vert \nonumber\\
&  =\sum_{k,l=1}^{N}\left(  \sum_{h=1}^{N}V_{hk}\left(  t\right)  V_{hl}%
^{\ast}\left(  t\right)  \right)  \left\vert s_{k}\left(  t\right)
\right\rangle \left\langle s_{l}\left(  t\right)  \right\vert \nonumber\\
&  =\sum_{k,l=1}^{N}\left\vert s_{k}\left(  t\right)  \right\rangle
\left\langle s_{l}\left(  t\right)  \right\vert \delta_{kl}\nonumber\\
&  =\sum_{k=1}^{N}\left\vert s_{k}\left(  t\right)  \right\rangle \left\langle
s_{k}\left(  t\right)  \right\vert \nonumber\\
&  =%
{\displaystyle\sum\limits_{k=1}^{N}}
p_{k}\left(  t\right)  \left\vert n_{k}(t)\right\rangle \left\langle
n_{k}(t)\right\vert \nonumber\\
&  =\rho\left(  t\right)
\end{align}
that is, $\rho\left(  t\right)  $ can be generally decomposed as%
\begin{equation}
\rho\left(  t\right)  =\sum_{h=1}^{N}\left\vert u_{h}\left(  t\right)
\right\rangle \left\langle u_{h}\left(  t\right)  \right\vert \text{.}
\label{cosmo9}%
\end{equation}
Let us consider now two neighboring nondegenerate states $\rho\left(
t\right)  $ and $\rho\left(  t+dt\right)  $ specified by the following
ensembles of pure states,%
\begin{equation}
\rho\left(  t\right)  \overset{\text{def}}{=}\left\{
{\displaystyle\sum\limits_{k=1}^{N}}
V_{hk}\left(  t\right)  \sqrt{p_{k}\left(  t\right)  }\left\vert n_{k}\left(
t\right)  \right\rangle \right\}  =\left\{  \left\vert u_{h}\left(  t\right)
\right\rangle \right\}  \label{cosmo10}%
\end{equation}
and,%
\begin{equation}
\rho\left(  t+dt\right)  \overset{\text{def}}{=}\left\{
{\displaystyle\sum\limits_{k=1}^{N}}
V_{hk}\left(  t+dt\right)  \sqrt{p_{k}\left(  t+dt\right)  }\left\vert
n_{k}\left(  t+dt\right)  \right\rangle \right\}  =\left\{  \left\vert
u_{h}\left(  t+dt\right)  \right\rangle \right\}  \text{,} \label{cosmo11}%
\end{equation}
respectively. For completeness, we note that $V_{hk}\left(  t\right)
=\left\vert V_{hk}\left(  t\right)  \right\vert e^{i\arg\left[  V_{hk}\left(
t\right)  \right]  }\in%
\mathbb{C}
$ for any $1\leq h$, $k\leq N$. In particular, one recovers the original
construction proposed originally by Sj\"{o}qvist when%
\begin{equation}
V_{hk}\left(  t\right)  =\delta_{hk}e^{if_{k}\left(  t\right)  }\text{, and
}\left\vert u_{h}\left(  t\right)  \right\rangle =\sqrt{p_{h}\left(  t\right)
}e^{if_{h}\left(  t\right)  }\left\vert n_{h}(t)\right\rangle \text{.}%
\end{equation}
Using the decompositions in Eqs. (\ref{cosmo10}) and (\ref{cosmo11}), the
generalization $\tilde{d}_{\mathrm{Sj\ddot{o}qvist}}^{2}\left(  t\text{,
}t+dt\right)  $ of $d_{\mathrm{Sj\ddot{o}qvist}}^{2}\left(  t\text{,
}t+dt\right)  $ in Eq. (\ref{cosmo3}) becomes%
\begin{equation}
\tilde{d}_{\mathrm{Sj\ddot{o}qvist}}^{2}\left(  t\text{, }t+dt\right)
\overset{\text{def}}{=}\underset{\left\{  V\left(  t\right)  \text{, }V\left(
t+dt\right)  \right\}  }{\min}%
{\displaystyle\sum\limits_{h=1}^{N}}
\left\Vert \left\vert u_{h}\left(  t\right)  \right\rangle -\left\vert
u_{h}\left(  t+dt\right)  \right\rangle \right\Vert ^{2}\text{,} \label{cos12}%
\end{equation}
that is,%
\begin{equation}
\tilde{d}_{\mathrm{Sj\ddot{o}qvist}}^{2}\left(  t\text{, }t+dt\right)  =%
{\displaystyle\sum\limits_{h=1}^{N}}
\left\Vert
{\displaystyle\sum\limits_{k=1}^{N}}
V_{hk}\left(  t\right)  \sqrt{p_{k}\left(  t\right)  }\left\vert n_{k}\left(
t\right)  \right\rangle -%
{\displaystyle\sum\limits_{k=1}^{N}}
V_{hk}\left(  t+dt\right)  \sqrt{p_{k}\left(  t+dt\right)  }\left\vert
n_{k}\left(  t+dt\right)  \right\rangle \right\Vert ^{2}\text{.} \label{cos13}%
\end{equation}
To obtain a more compact expression of $\tilde{d}_{\mathrm{Sj\ddot{o}qvist}%
}^{2}\left(  t\text{, }t+dt\right)  $, we note that%
\begin{align}%
{\displaystyle\sum\limits_{h=1}^{N}}
\left\Vert \left\vert u_{h}\left(  t\right)  \right\rangle -\left\vert
u_{h}\left(  t+dt\right)  \right\rangle \right\Vert ^{2}  &
=2-2\operatorname{Re}\left[  \sum_{h=1}^{N}\left\langle u_{h}\left(  t\right)
\left\vert u_{h}\left(  t+dt\right)  \right.  \right\rangle \right]
\nonumber\\
&  =2-2\operatorname{Re}\left[  \sum_{h,k,k^{\prime}}V_{hk}^{\ast}\left(
t\right)  \left\langle s_{k}\left(  t\right)  \left\vert s_{k^{\prime}}\left(
t+dt\right)  \right.  \right\rangle V_{hk^{\prime}}\left(  t+dt\right)
\right] \nonumber\\
&  =2-2\operatorname{Re}\left[  \sum_{h,k,k^{\prime}}S_{kk^{\prime}%
}V_{hk^{\prime}}\left(  t+dt\right)  V_{hk}^{\ast}\left(  t\right)  \right]
\nonumber\\
&  =2-2\operatorname{Re}\mathrm{tr}\left[  S_{t}\left(  dt\right)  V\left(
t+dt\right)  V^{\dagger}\left(  t\right)  \right]  \text{,}%
\end{align}
that is,%
\begin{equation}%
{\displaystyle\sum\limits_{h=1}^{N}}
\left\Vert \left\vert u_{h}\left(  t\right)  \right\rangle -\left\vert
u_{h}\left(  t+dt\right)  \right\rangle \right\Vert ^{2}=2-2\operatorname{Re}%
\mathrm{tr}\left[  S_{t}(dt)V\left(  t+dt\right)  V^{\dagger}\left(  t\right)
\right]  \text{.} \label{cosmo14}%
\end{equation}
The matrix $S_{t}\left(  dt\right)  $ in Eq. (\ref{cosmo14}) is an overlap
matrix with coefficients $S_{kk^{\prime}}$ defined as%
\begin{equation}
S_{kk^{\prime}}\overset{\text{def}}{=}\left\langle s_{k}\left(  t\right)
\left\vert s_{k^{\prime}}\left(  t+dt\right)  \right.  \right\rangle
=\sqrt{p_{k}\left(  t\right)  p_{k^{\prime}}\left(  t+dt\right)  }\left\langle
n_{k}\left(  t\right)  \left\vert n_{k^{\prime}}\left(  t+dt\right)  \right.
\right\rangle \text{.}%
\end{equation}
Combining Eqs. (\ref{cos12}) and (\ref{cosmo14}), we finally get%
\begin{equation}
\tilde{d}_{\mathrm{Sj\ddot{o}qvist}}^{2}\left(  t\text{, }t+dt\right)
=\underset{\left\{  V\left(  t\right)  \text{, }V\left(  t+dt\right)
\right\}  }{\min}\left\{  2-2\operatorname{Re}\mathrm{tr}\left[
S_{t}(dt)V\left(  t+dt\right)  V^{\dagger}\left(  t\right)  \right]  \right\}
\text{.} \label{cosmo15}%
\end{equation}
In what follows, we shall see the emergence of the Bures metric by explicitly
evaluating the \emph{minimum} that specifies $\tilde{d}_{\mathrm{Sj\ddot
{o}qvist}}^{2}\left(  t\text{, }t+dt\right)  $ in Eq. (\ref{cosmo15}).

\subsection{Emergence of the Bures metric}

We begin by observing that the polar decomposition of $S_{t}(dt)$ is given by
\cite{nielsen00},
\begin{equation}
S_{t}(dt)=\left\vert S_{t}(dt)\right\vert U_{t}\left(  dt\right)  \text{,}%
\end{equation}
where $\left\vert S_{t}(dt)\right\vert \overset{\text{def}}{=}\sqrt
{S_{t}(dt)S_{t}^{\dagger}(dt)}$ and $U_{t}\left(  dt\right)  $ is a unitary
matrix. Then, we note that minimizing $2-2\operatorname{Re}\mathrm{tr}\left[
S_{t}(dt)V\left(  t+dt\right)  V^{\dagger}\left(  t\right)  \right]  $ is
equivalent to maximizing $2\operatorname{Re}\mathrm{tr}\left[  S_{t}%
(dt)V\left(  t+dt\right)  V^{\dagger}\left(  t\right)  \right]  $ with respect
to $\left\{  V\left(  t\right)  \text{, }V\left(  t+dt\right)  \right\}  $.
Furthermore, we make two remarks. First of all, $\operatorname{Re}\left(
z\right)  \leq\left\vert z\right\vert $ for any $z\in%
\mathbb{C}
$. Second of all, \textrm{tr}$\left\vert A\right\vert \geq\left\vert
\mathrm{tr}\left(  AU_{A}\right)  \right\vert $ for any operator $A$ and
unitary $U_{A}$ with $\underset{U_{A}}{\max}\left\vert \mathrm{tr}\left(
AU_{A}\right)  \right\vert =$ \textrm{tr}$\left\vert A\right\vert $ obtained
by choosing $U_{A}=V_{A}^{\dagger}$ where $A=\left\vert A\right\vert V_{A}$ is
the polar decomposition of $A$ \cite{nielsen00,wilde13}. Given this set of
preliminary observations, we have that%
\begin{align}
\operatorname{Re}\mathrm{tr}\left[  S_{t}(dt)V\left(  t+dt\right)  V^{\dagger
}\left(  t\right)  \right]   &  =\operatorname{Re}\mathrm{tr}\left[
\left\vert S_{t}(dt)\right\vert U_{t}\left(  dt\right)  V\left(  t+dt\right)
V^{\dagger}\left(  t\right)  \right] \nonumber\\
&  \leq\left\vert \mathrm{tr}\left[  \left\vert S_{t}(dt)\right\vert
U_{t}\left(  dt\right)  V\left(  t+dt\right)  V^{\dagger}\left(  t\right)
\right]  \right\vert \nonumber\\
&  \leq\mathrm{tr}\left\vert S_{t}(dt)\right\vert \text{,}%
\end{align}
that is,%
\begin{equation}
\underset{\left\{  V\left(  t\right)  \text{, }V\left(  t+dt\right)  \right\}
}{\max}\operatorname{Re}\mathrm{tr}\left[  S_{t}(dt)V\left(  t+dt\right)
V^{\dagger}\left(  t\right)  \right]  =\mathrm{tr}\left\vert S_{t}%
(dt)\right\vert \label{cosmo16}%
\end{equation}
is obtained by choosing $\left\{  V\left(  t\right)  \text{, }V\left(
t+dt\right)  \right\}  $ such that the following condition is satisfied,%
\begin{equation}
U_{t}\left(  dt\right)  V\left(  t+dt\right)  V^{\dagger}\left(  t\right)
=I\text{.} \label{condition2}%
\end{equation}
Interestingly, we point out that the maximization procedure in Eq.
(\ref{cosmo16}) is similar to the use of the variational characterization of
the trace norm that one employs to prove Uhlmann's theorem (see, for instance,
Lemma $9.5$ in Ref. \cite{nielsen00} and Property $9.1.6$ in Ref.
\cite{wilde13}). We also remark\textbf{ }that Eq. (\ref{condition2}) is a
constraint equation that can be regarded as the operator-analogue of the
parallel transport condition in Eq. (\ref{condo2}). For more details on this
point, we refer to Appendix A. Finally, employing Eqs. (\ref{cosmo15}) and
(\ref{cosmo16}), we get%
\begin{equation}
\tilde{d}_{\mathrm{Sj\ddot{o}qvist}}^{2}\left(  t\text{, }t+dt\right)
=2-2\mathrm{tr}\left\vert S_{t}(dt)\right\vert \text{.} \label{cosmo17}%
\end{equation}
We shall now show that $\tilde{d}_{\mathrm{Sj\ddot{o}qvist}}^{2}\left(
t\text{, }t+dt\right)  $ in Eq. (\ref{cosmo17}) is indeed the Bures metric
$d_{\mathrm{Bures}}^{2}\left(  t\text{, }t+dt\right)  $ defined as
\cite{nielsen00,karol06},
\begin{equation}
d_{\mathrm{Bures}}^{2}\left(  t\text{, }t+dt\right)  \overset{\text{def}%
}{=}2-2\mathrm{tr}\left[  \sqrt{\rho^{1/2}\left(  t\right)  \rho\left(
t+dt\right)  \rho^{1/2}\left(  t\right)  }\right]  \text{.} \label{burino}%
\end{equation}
Observe that $\left\vert S_{t}(dt)\right\vert ^{2}=S_{t}(dt)S_{t}^{\dagger
}(dt)$, where%
\begin{equation}
\left[  S_{t}(dt)S_{t}^{\dagger}(dt)\right]  _{kk^{\prime\prime}}=%
{\displaystyle\sum\limits_{k^{\prime}=1}^{N}}
\left\langle s_{k}\left(  t\right)  \left\vert s_{k^{\prime}}\left(
t+dt\right)  \right.  \right\rangle \left\langle s_{k^{\prime}}\left(
t+dt\right)  \left\vert s_{k^{\prime\prime}}\left(  t\right)  \right.
\right\rangle \text{.}%
\end{equation}
After some algebra, we note that $\rho^{1/2}\left(  t\right)  \rho\left(
t+dt\right)  \rho^{1/2}\left(  t\right)  =\left\vert S_{t}(dt)\right\vert
^{2}$. Indeed, we have%
\begin{align}
\rho^{1/2}\left(  t\right)  \rho\left(  t+dt\right)  \rho^{1/2}\left(
t\right)   &  =\left(
{\displaystyle\sum\limits_{k=1}^{N}}
\sqrt{p_{k}\left(  t\right)  }\left\vert n_{k}(t)\right\rangle \left\langle
n_{k}(t)\right\vert \right)  \left(
{\displaystyle\sum\limits_{k^{\prime}=1}^{N}}
p_{k^{\prime}}\left(  t+dt\right)  \left\vert n_{k^{\prime}}%
(t+dt)\right\rangle \left\langle n_{k^{\prime}}(t+dt)\right\vert \right)
\nonumber\\
&  \left(
{\displaystyle\sum\limits_{k^{\prime\prime}=1}^{N}}
\sqrt{p_{k^{\prime\prime}}\left(  t\right)  }\left\vert n_{k^{\prime\prime}%
}(t)\right\rangle \left\langle n_{k^{\prime\prime}}(t)\right\vert \right)
\nonumber\\
&  =%
{\displaystyle\sum\limits_{k,k^{\prime},k^{\prime\prime}=1}^{N}}
\left[
\begin{array}
[c]{c}%
\left\vert n_{k}(t)\right\rangle \left(  \sqrt{p_{k}\left(  t\right)
p_{k^{\prime}}\left(  t+dt\right)  }\left\langle n_{k}(t)\left\vert
n_{k^{\prime}}(t+dt)\right.  \right\rangle \right) \\
\left(  \sqrt{p_{k^{\prime}}\left(  t+dt\right)  p_{k^{\prime\prime}}\left(
t\right)  }\left\langle n_{k^{\prime}}(t+dt)\left\vert n_{k^{\prime\prime}%
}(t)\right.  \right\rangle \right)  \left\langle n_{k^{\prime\prime}%
}(t)\right\vert
\end{array}
\right] \nonumber\\
&  =%
{\displaystyle\sum\limits_{k,k^{\prime},k^{\prime\prime}=1}^{N}}
\left\vert n_{k}(t)\right\rangle \left(  \left\langle s_{k}\left(  t\right)
\left\vert s_{k^{\prime}}\left(  t+dt\right)  \right.  \right\rangle \right)
\left(  \left\langle s_{k^{\prime}}\left(  t+dt\right)  \left\vert
s_{k^{\prime\prime}}\left(  t\right)  \right.  \right\rangle \right)
\left\langle n_{k^{\prime\prime}}(t)\right\vert \nonumber\\
&  =%
{\displaystyle\sum\limits_{k,k^{\prime\prime}=1}^{N}}
\left\vert n_{k}(t)\right\rangle \left[  \sum_{k=1}^{N}\left\langle
s_{k}\left(  t\right)  \left\vert s_{k^{\prime}}\left(  t+dt\right)  \right.
\right\rangle \left\langle s_{k^{\prime}}\left(  t+dt\right)  \left\vert
s_{k^{\prime\prime}}\left(  t\right)  \right.  \right\rangle \right]
\left\langle n_{k^{\prime\prime}}(t)\right\vert \nonumber\\
&  =%
{\displaystyle\sum\limits_{k,k^{\prime\prime}=1}^{N}}
\left\vert n_{k}(t)\right\rangle \left[  S_{t}\left(  dt\right)
S_{t}^{\dagger}\left(  dt\right)  \right]  _{kk^{\prime\prime}}\left\langle
n_{k^{\prime\prime}}(t)\right\vert \nonumber\\
&  =%
{\displaystyle\sum\limits_{k,k^{\prime\prime}=1}^{N}}
\left[  S_{t}\left(  dt\right)  S_{t}^{\dagger}\left(  dt\right)  \right]
_{kk^{\prime\prime}}\left\vert n_{k}(t)\right\rangle \left\langle
n_{k^{\prime\prime}}(t)\right\vert \nonumber\\
&  =S_{t}\left(  dt\right)  S_{t}^{\dagger}\left(  dt\right) \nonumber\\
&  =\left\vert S_{t}(dt)\right\vert ^{2}\text{.}%
\end{align}
In conclusion, we arrive at the following relations%
\begin{equation}
d_{\mathrm{Bures}}^{2}\left(  t\text{, }t+dt\right)  =\tilde{d}%
_{\mathrm{Sj\ddot{o}qvist}}^{2}\left(  t\text{, }t+dt\right)  \neq
d_{\mathrm{Sj\ddot{o}qvist}}^{2}\left(  t\text{, }t+dt\right)  \text{.}%
\end{equation}
More specifically, we have $\tilde{d}_{\mathrm{Sj\ddot{o}qvist}}^{2}\left(
t\text{, }t+dt\right)  \leq d_{\mathrm{Sj\ddot{o}qvist}}^{2}\left(  t\text{,
}t+dt\right)  $ since the minimization procedure that specifies $\tilde
{d}_{\mathrm{Sj\ddot{o}qvist}}^{2}\left(  t\text{, }t+dt\right)  $ is extended
to arbitrary unitary $\left\{  V\left(  t\right)  \text{, }V\left(
t+dt\right)  \right\}  $ while, instead, the minimization procedure that
specifies $d_{\mathrm{Sj\ddot{o}qvist}}^{2}\left(  t\text{, }t+dt\right)  $ is
limited to unitary matrices of the form $\left\{  V\left(  t\right)  \text{,
}V\left(  t+dt\right)  \right\}  $ with $V_{hk}\left(  t\right)  =\delta
_{hk}e^{if_{k}\left(  t\right)  }$. \ With this last remark, we end our
mathematical discussion on the emergence of the Bures metric from a
generalized version of the Sj\"{o}qvist original metric construction. However,
to better grasp the physical differences between the Sj\"{o}qvist and Bures
metrics, we focus on thermal mixed states in the next section.

\section{ Sj\"{o}qvist and Bures metrics for thermal states}

In this section, we cast the Sj\"{o}qvist and the Bures metrics in two forms
that are especially convenient for an insightful geometric comparison. In
particular, we illustrate this comparison with an explicit example in which
the physical system is specified by a spin-qubit in an arbitrarily oriented
uniform and stationary magnetic field in thermal equilibrium with a
finite-temperature reservoir.

\subsection{Suitable recast of metrics}

We begin by observing that, in the Sj\"{o}qvist case (see Eq. (\ref{cosmo4B}%
)), the metric (infinitesimal line element) can be decomposed in terms of a
classical and a nonclassical contribution,%
\begin{equation}
ds_{\mathrm{Sj\ddot{o}qvist}}^{2}=\left(  ds_{\mathrm{Sj\ddot{o}qvist}}%
^{2}\right)  ^{\mathrm{c}}+\left(  ds_{\mathrm{Sj\ddot{o}qvist}}^{2}\right)
^{\mathrm{nc}}\text{.}%
\end{equation}

It happens that $\left(  ds_{\mathrm{Sj\ddot{o}qvist}}^{2}\right)
^{\mathrm{c}}$ and $\left(  ds_{\mathrm{Sj\ddot{o}qvist}}^{2}\right)
^{\mathrm{nc}}$ can be conveniently written as \cite{erik20},
\begin{equation}
\left(  ds_{\mathrm{Sj\ddot{o}qvist}}^{2}\right)  ^{\mathrm{c}}%
\overset{\text{def}}{=}\frac{1}{4}\sum_{n}\frac{dp_{n}^{2}}{p_{n}}\text{, and
}\left(  ds_{\mathrm{Sj\ddot{o}qvist}}^{2}\right)  ^{\mathrm{nc}%
}\overset{\text{def}}{=}\sum_{n}p_{n}\left\langle dn|(\mathrm{I}-\left\vert
n\right\rangle \left\langle n\right\vert )|dn\right\rangle \text{,}
\label{snc}%
\end{equation}
respectively. To recast $\left(  ds_{\mathrm{Sj\ddot{o}qvist}}^{2}\right)
^{\mathrm{nc}}$ in Eq. (\ref{snc}) in a suitable manner for thermal states
$\rho\overset{\text{def}}{=}\sum_{n}p_{n}\left\vert n\right\rangle
\left\langle n\right\vert $ where $\left\{  \left\vert n\right\rangle
\right\}  $ denotes the eigenbasis of $\rho$ with eigenvalues $\left\{
p_{n}\right\}  $ and $1\leq n\leq N$, we note that%
\begin{equation}
\left\langle dn\left\vert dn\right.  \right\rangle =\left\langle dn\left\vert
n\right.  \right\rangle \left\langle n\left\vert dn\right.  \right\rangle
+\sum_{k\text{, }k\neq n}\left\langle dn\left\vert k\right.  \right\rangle
\left\langle k\left\vert dn\right.  \right\rangle \text{.} \label{from1}%
\end{equation}
Furthermore, assuming that the Hamiltonian operator \textrm{H }satisfies the
relation \textrm{H}$\left\vert n\right\rangle =E_{n}\left\vert n\right\rangle
$ with $\left\{  E_{n}\right\}  $ and $\left\{  \left\vert n\right\rangle
\right\}  $ being eigenvalues and eigenvectors of \textrm{H}, respectively, we
find after some clever algebraic manipulations that%
\begin{equation}
\left\langle k\left\vert dn\right.  \right\rangle \left\langle dn\left\vert
k\right.  \right\rangle =\left\vert \frac{\left\langle k\left\vert
d\mathrm{H}\right\vert n\right\rangle }{E_{n}-E_{k}}\right\vert ^{2}\text{.}
\label{from2}%
\end{equation}
Then, exploiting Eqs. (\ref{from1}) and (\ref{from2}), $\left(
ds_{\mathrm{Sj\ddot{o}qvist}}^{2}\right)  ^{\mathrm{nc}}$ in Eq. (\ref{snc})
can be finally expressed as%
\begin{equation}
\left(  ds_{\mathrm{Sj\ddot{o}qvist}}^{2}\right)  ^{\mathrm{nc}}=\sum_{n\neq
k}\frac{e^{-\beta E_{n}}+e^{-\beta E_{k}}}{2\mathcal{Z}}\left\vert
\frac{\left\langle n|dH|k\right\rangle }{E_{n}-E_{k}}\right\vert ^{2}\text{.}
\label{from3}%
\end{equation}
In Eq. (\ref{from3}), $\mathcal{Z}\overset{\text{def}}{\mathcal{=}}%
\mathrm{tr}\left(  e^{-\beta\mathrm{H}}\right)  $ is the partition function of
the system, $p_{n}\overset{\text{def}}{\mathcal{=}}e^{-\beta E_{n}%
}/\mathcal{Z}$, $\beta\overset{\text{def}}{\mathcal{=}}\left(  k_{B}T\right)
^{-1}$, and $k_{B}$ is the Boltzmann constant. Eq. (\ref{from3}) is an
interesting result of our work and denotes the suitable recast of $\left(
ds_{\mathrm{Sj\ddot{o}qvist}}^{2}\right)  ^{\mathrm{nc}}$ for thermal quantum
states we were looking for. We need to find now the analog of Eq.
(\ref{from3}) for the Bures case.

In the Bures case, the metric (infinitesimal line element) can be decomposed
in terms of a classical and a nonclassical contribution,%
\begin{equation}
ds_{\mathrm{Bures}}^{2}=\left(  ds_{\mathrm{Bures}}^{2}\right)  ^{\mathrm{c}%
}+\left(  ds_{\mathrm{Bures}}^{2}\right)  ^{\mathrm{nc}}\text{.}
\label{stefano}%
\end{equation}
Focusing on thermal quantum states $\rho\overset{\text{def}}{=}\sum_{n}%
p_{n}\left\vert n\right\rangle \left\langle n\right\vert $ as pointed out
earlier, it can be shown that $\left(  ds_{\mathrm{Bures}}^{2}\right)
^{\mathrm{c}}=$ $\left(  ds_{\mathrm{Sj\ddot{o}qvist}}^{2}\right)
^{\mathrm{c}}$ in Eq. (\ref{snc}) and $\left(  ds_{\mathrm{Bures}}^{2}\right)
^{\mathrm{nc}}$ can be expressed as \cite{zanardi07B,hubner92},
\begin{equation}
\left(  ds_{\mathrm{Bures}}^{2}\right)  ^{\mathrm{nc}}=\sum_{n\neq k}%
\frac{e^{-\beta E_{n}}+e^{-\beta E_{k}}}{2\mathcal{Z}}\left(  \frac{e^{-\beta
E_{n}}-e^{-\beta E_{k}}}{e^{-\beta E_{n}}+e^{-\beta E_{k}}}\right)
^{2}\left\vert \frac{\left\langle n|dH|k\right\rangle }{E_{n}-E_{k}%
}\right\vert ^{2}\text{.} \label{from4}%
\end{equation}
For completeness, note that a general expression of $ds_{\mathrm{Bures}}^{2}$
in Eq. (\ref{stefano}) can be obtained by replacing $e^{-\beta E_{n}%
}/\mathcal{Z}$ with an arbitrary $p_{n}$ as remarked in Ref. \cite{zanardi07B}%
. We observe that Eq. (\ref{from4}) is, modulo a clever rewriting that suits
our comparative discussion between the Sj\"{o}qvist and the Bures metrics
here, equivalent to Eq. (6) in Ref. \cite{zanardi07B}. The difference between
the Sj\"{o}qvist and the Bures metrics $ds_{\mathrm{Sj\ddot{o}qvist}}^{2}$ and
$ds_{\mathrm{Bures}}^{2}$ appears in their non-classical metric components
$g_{\mathrm{Sj\ddot{o}qvist}}^{nc}$ and $g_{\mathrm{Bures}}^{nc}$. In
particular, focusing on Eqs. (\ref{from3}) and (\ref{from4}), the difference
between these components, in turn, tends to vanish when the minimum separation
between the modulus of two distinct quantum-mechanical energy levels $E_{n}$
and $E_{k}$ of the system is much greater than the characteristic thermal
energy $k_{B}T$, i.e.%
\begin{equation}
\underset{n\neq k}{\min}\left\vert E_{n}-E_{k}\right\vert \gg k_{B}T\text{.}
\label{relazione}%
\end{equation}
Clearly, Eq. (\ref{relazione}) is satisfied when the temperature $T$
approaches zero (i.e., asymptotic limit of $\beta$ approaching infinity) with
$\left\vert E_{n}-E_{k}\right\vert $ finite (and nonzero) for any $n\neq k$.
In this case, mixed quantum states tend to become pure states and, in
particular, both metrics (i.e., Sj\"{o}qvist and Bures) reduce to the
Fubini-Study metric.

\subsection{Illustrative example}

To better grasp the difference between these two metrics as reported in Eqs.
(\ref{from3}) and (\ref{from4}), we discuss an explicit example. Let us take
into consideration a spin-$1/2$ particle specified by an electron of $m$,
charge $-e$ with $e\geq0$ immersed in an external magnetic field $\vec
{B}\left(  t\right)  $. The Hamiltonian of this system can be
quantum-mechanically specified by the Hermitian operator $\mathrm{H}\left(
t\right)  \overset{\text{def}}{=}-\vec{\mu}\mathbf{\cdot}\vec{B}\left(
t\right)  $, where $\vec{\mu}\overset{\text{def}}{=}-\left(  e/m\right)
\vec{s}$ is the electron magnetic moment operator and $\vec{s}%
\overset{\text{def}}{=}\left(  \hslash/2\right)  \vec{\sigma}$ is the spin
operator. Naturally, $\hslash\overset{\text{def}}{=}h/(2\pi)$ denotes the
reduced Planck constant and $\vec{\sigma}\overset{\text{def}}{=}\left(
\sigma_{x}\text{, }\sigma_{y}\text{, }\sigma_{z}\right)  $ represents the
Pauli spin vector operator. If we consider a time-independent, uniform, and
arbitrarily magnetic field given by $\vec{B}\overset{\text{def}}{=}B_{x}%
\hat{x}+B_{y}\hat{y}+B_{z}\hat{z}$ and introduce the frequency vector
$\vec{\omega}\overset{\text{def}}{=}\left(  \omega_{x}\text{, }\omega
_{y}\text{, }\omega_{z}\right)  =\left(  (e/m)B_{x}\text{, }(e/m)B_{y}\text{,
}(e/m)B_{z}\right)  $, the spin-$1/2$ qubit (SQ) Hamiltonian becomes%
\begin{equation}
\mathrm{H}_{\mathrm{SQ}}\left(  \vec{\omega}\right)  \overset{\text{def}%
}{=}\frac{\hslash}{2}\left(  \vec{\omega}\cdot\vec{\sigma}\right)  \text{.}
\label{ham}%
\end{equation}
Note that with the sign convention used for\textbf{ }$\mathrm{H}_{\mathrm{SQ}%
}\left(  \vec{\omega}\right)  $ in Eq. (\ref{ham}), when\textbf{ }$\vec
{\omega}=\omega_{z}\hat{z}$\textbf{ }with\textbf{ }$\omega_{z}>0$\textbf{, }we
have that\textbf{ }$\left\vert 1\right\rangle $\textbf{ (}$\left\vert
0\right\rangle $\textbf{) }represents the ground (excited) state of the system
with energy\textbf{ }$-\hslash\omega_{z}/2$\textbf{ (}$+\hslash\omega_{z}%
/2$\textbf{).} Furthermore, let us suppose that the system specified by the
Hamiltonian $\mathrm{H}_{\mathrm{SQ}}$ in Eq. (\ref{ham}) is in thermal
equilibrium with a reservoir at non-zero temperature $T$. Then, quantum
statistical mechanics \cite{huang87} specifies that the system has temperature
$T$ and its state is characterized by a thermal state \cite{strocchi08}
specified by a density matrix $\rho$ given by%
\begin{equation}
\rho_{\mathrm{SQ}}\left(  \beta\text{, }\vec{\omega}\right)
\overset{\text{def}}{=}\frac{e^{-\beta\mathrm{H}_{\mathrm{SQ}}\left(
\vec{\omega}\right)  }}{\mathrm{tr}\left(  e^{-\beta\mathrm{H}_{\mathrm{SQ}%
}\left(  \vec{\omega}\right)  }\right)  }\text{.} \label{den}%
\end{equation}
In Eq. (\ref{den}), $\beta\overset{\text{def}}{=}\left(  k_{B}T\right)  ^{-1}$
is the so-called inverse temperature, while $k_{B}$ denotes the Boltzmann
constant. Using Eqs. (\ref{ham}) and (\ref{den}), one obtains after some
algebra that the formal expression of the thermal state $\rho_{\mathrm{SQ}%
}\left(  \beta\text{, }\vec{\omega}\right)  $ is given by%
\begin{equation}
\rho_{\mathrm{SQ}}\left(  \beta\text{, }\vec{\omega}\right)  =\frac{1}%
{2}\left[  \mathrm{I}-\tanh\left(  \beta\frac{\hslash\omega}{2}\right)
\frac{\vec{\omega}\cdot\vec{\sigma}}{\omega}\right]  \text{,} \label{den1}%
\end{equation}
with $\omega\overset{\text{def}}{=}\sqrt{\omega_{x}^{2}+\omega_{y}^{2}%
+\omega_{z}^{2}}$ denoting the magnitude of the frequency vector $\vec{\omega
}$ and $\mathrm{I}$ in Eq. (\ref{den1}) being the identify operator. Finally,
assuming to keep $\omega_{x}$-fixed$\neq0$, $\omega_{y}$-fixed$\neq0$ and, at
the same time, tuning only the two parameters $\beta$ and $\omega_{z}$, the
Sj\"{o}qvist and the Bures metrics specifying the distance between the two
neighboring mixed states $\rho_{\mathrm{SQ}}$ and $\rho_{\mathrm{SQ}}%
+d\rho_{\mathrm{SQ}}$ can be analytically shown to be equal to%
\begin{equation}
g_{ij}^{\mathrm{Sj\ddot{o}qvist}}\left(  \beta\text{, }\omega_{z}\right)
=\frac{\hslash^{2}}{16}\left[  1-\tanh^{2}\left(  \beta\frac{\hslash\omega}%
{2}\right)  \right]  \left(
\begin{array}
[c]{cc}%
\omega^{2} & \beta\omega_{z}\\
\beta\omega_{z} & \beta^{2}\left(  \frac{\omega_{z}}{\omega}\right)
^{2}+\frac{4}{\hslash^{2}}\frac{\omega_{x}^{2}+\omega_{y}^{2}}{\omega^{4}%
}\frac{1}{1-\tanh^{2}\left(  \beta\frac{\hslash\omega}{2}\right)  }%
\end{array}
\right)  \text{,} \label{s1}%
\end{equation}
and%
\begin{equation}
g_{ij}^{\mathrm{Bures}}\left(  \beta\text{, }\omega_{z}\right)  =\frac
{\hslash^{2}}{16}\left[  1-\tanh^{2}\left(  \beta\frac{\hslash\omega}%
{2}\right)  \right]  \left(
\begin{array}
[c]{cc}%
\omega^{2} & \beta\omega_{z}\\
\beta\omega_{z} & \beta^{2}\left(  \frac{\omega_{z}}{\omega}\right)
^{2}+\frac{4}{\hslash^{2}}\frac{\omega_{x}^{2}+\omega_{y}^{2}}{\omega^{4}%
}\frac{\tanh^{2}\left(  \beta\frac{\hslash\omega}{2}\right)  }{1-\tanh
^{2}\left(  \beta\frac{\hslash\omega}{2}\right)  }%
\end{array}
\right)  \text{,} \label{s2}%
\end{equation}
respectively, with $1\leq i$, $j\leq2$ (where $1\leftrightarrow\beta$ and
$2\leftrightarrow\omega_{z}$). For explicit technical details on how to
analytically calculate the Sj\"{o}qvist and the Bures metrics, we refer to
Ref. \cite{cafaroepj}. From Eqs. (\ref{s1}) and (\ref{s2}), it is clear that
the Sj\"{o}qvist and the Bures metrics only differ in the non-classical
contribution $\left[  g_{22}\left(  \beta\text{, }\omega_{z}\right)  \right]
_{\mathrm{nc}}$ of their $g_{22}\left(  \beta\text{, }\omega_{z}\right)  $
metric component.\ Specifically, we observe that%
\begin{equation}
0\leq\left[  g_{22}^{\mathrm{Bures}}\left(  \beta\text{, }\omega_{z}\right)
\right]  _{\mathrm{nc}}/\left[  g_{22}^{\mathrm{Sj\ddot{o}qvist}}\left(
\beta\text{, }\omega_{z}\right)  \right]  _{\mathrm{nc}}=\tanh^{2}\left(
\beta\frac{\hslash\omega}{2}\right)  \leq1\text{.} \label{s3}%
\end{equation}
Interestingly, for a two-level system with $E_{1}=\hslash\omega/2$ and
$E_{2}=-\hslash\omega/2$, the factor $\left[  \left(  e^{-\beta E_{1}%
}-e^{-\beta E_{2}}\right)  /\left(  e^{-\beta E_{1}}+e^{-\beta E_{2}}\right)
\right]  ^{2}$ in Eq. (\ref{from4}) becomes exactly $\tanh^{2}\left[
\beta\left(  \hslash\omega/2\right)  \right]  $ (i.e., the ratio in Eq.
(\ref{s3})).

We note that, in the limiting case in which $\vec{\omega}=\left(  0\text{,
}0\text{, }\omega_{z}\right)  $, setting $k_{B}=1$, $\beta=t^{-1}$, and
$\omega_{z}=t$, our Eq. (\ref{s2}) reduces to the last relation found by
Zanardi and collaborators in Ref. \cite{zanardi07B}. In Ref. \cite{zanardi07B}%
, the limiting scenario considered by Zanardi and collaborators corresponds to
the case of a one-dimensional quantum Ising model in a transverse magnetic
field $h\equiv B_{z}$ with $\left\vert h\right\vert \gg1$. When $\left\vert
h\right\vert \gg1$, the lowest order approximation of the quantum Ising
Hamiltonian is \textrm{H}$=h\sum_{i}\sigma_{i}^{z}$. In this approximation,
the Bures metric between two neighboring thermal states parametrized by
$\left\{  \beta\text{, }h\right\}  $ and emerging from this approximated
Hamiltonian vanishes. In our analysis, the degeneracy of the Bures metric
appears when the spin-qubit is immersed in a magnetic field oriented along the
$z$-axis. In particular, the metric has in this case only one nonvanishing
eigenvalue, its determinant vanishes, and no definition of connection and
curvature exists. In summary, no Riemannian structure survives at all when the
metric is degenerate. In Ref. \cite{zanardi07B}, the degeneration of the
metric can be removed by considering higher order approximations of the
quantum Ising Hamiltonian. In our case, instead, the degeneracy of the Bures
metric can be removed by considering more general orientations of the external
magnetic field. Interestingly, the degenerate scenario can be given a clear
interpretation, despite the absence of any Riemannian structure. Indeed, given
that the eigenvectors of the Bures metric tensor define the directions of
maximal and minimal growth of the line element $ds_{\mathrm{Bures}}^{2}$
\cite{zanardi07B}, the eigenvector of the metric related to the highest
eigenvalue defines at each point of the two-dimensional parametric\ plane the
direction along which the Uhlmann fidelity between two nearby states decreases
most quickly, i. e., the direction of highest distinguishability between two
neighboring thermal states. Therefore, when proceeding along the direction
specified by an eigenvector corresponding to the vanishing eigenvalue, one can
conclude that no change in the state of the system takes place.

For completeness, we reiterate that \ in this paper we limited our theoretical
discussions to nondegenerate density matrices for which Sj\"{o}qvist's
original metric is nonsingular. In particular, our explicit illustrative
example was specified by an Hamiltonian with nondegenerate eigenvalues
yielding nondegenerate density operators. However, degenerate thermal states
that emerge from degenerate-spectrum Hamiltonians are pervasive in physics
\cite{gubser05}. In these latter scenarios, insights on the physics of quantum
systems can be generally obtained by studying the geometry of thermal state
manifolds equipped with a generalized version of Sj\"{o}qvist's original
metric. This generalized metric is also suitable for degenerate mixed quantum
states and was proposed by Silva and collaborators in Ref. \cite{silva21}.

In conclusion, we point out that for pure quantum states $\left(  \rho
=\rho^{2}\right)  $ and for mixed quantum states $\left(  \rho\neq\rho
^{2}\right)  $ for which the non-commutative probabilistic structure
underlying quantum theory is invisible (i.e., in the classical scenario with
$\left[  \rho\text{, }\rho+d\rho\right]  =0$), the Bures and the Sj\"{o}qvist
metrics are essentially the same. Indeed, in the former and latter cases, they
reduce to the Fubini-Study and Fisher-Rao information metrics, respectively.
Instead, when considering mixed quantum states for which the non-commutative
probabilistic structure of quantum mechanics is visible (i.e., in the
non-classical scenario with $\left[  \rho\text{, }\rho+d\rho\right]  \neq0$),
the Bures and the Sj\"{o}qvist metrics are generally different. This latter
scenario has been explicitly illustrated in our proposed example.

In the next section, we shall investigate the monotonicity aspects of the
Sj\"{o}qvist metric for mixed states.

\section{Monotonicity of the Sj\"{o}qvist metric}

In this section, we discuss the monotonicity of the Sj\"{o}qvist metric in the
single-qubit case. Unlike the Bures metric, we shall see that the Sj\"{o}qvist
metric is not specified by a proper Morozova-Chentsov function and is not a
monotone metric. For some technical details on the monotonicity of the Bures
metric, see Appendix B.

\subsection{Preliminaries}

If a distance between classical probability distributions or quantum density
matrices expresses statistical distinguishability, then this distance must not
increase under coarse-graining. In particular, a metric that does not grow
under the action of a stochastic map is called monotone \cite{karol06}. In the
classical setting, the Fisher-Rao information metric is the unique
\cite{cencov81,campbell86}, except for a constant scale factor, Riemannian
metric that is invariant under Markov embeddings (i.e., stochastic maps). In
the quantum setting, instead, there are infinitely many monotone Riemannian
metrics on the space of quantum states \cite{karol06}. In the quantum case,
quantum stochastic maps are represented by completely positive and trace
preserving (CPTP) maps. If $D_{\mathrm{mon}}\left(  \rho\text{, }%
\sigma\right)  $ represents the distance between density matrices $\rho$ and
$\sigma$ that originates from a monotone metric, it must be%
\begin{equation}
D_{\mathrm{mon}}\left(  \Lambda\left(  \rho\right)  \text{, }\Lambda\left(
\sigma\right)  \right)  \leq D_{\mathrm{mon}}\left(  \rho\text{, }%
\sigma\right)  \text{,} \label{jovi}%
\end{equation}
for any CPTP map $\Lambda$. Morozova and Chentsov originally considered the
problem of finding monotone Riemannian metrics on the space of density
matrices \cite{morozova91}. However, although they proposed several
candidates, they did not present a single explicit example of a monotone
metric. It was Petz, building on the work of Morozova and Chentsov, who showed
the abundance of monotone metrics by exploiting the concept of operator
monotone function in Ref. \cite{petz96a}. A scalar function $f:I\rightarrow%
\mathbb{R}
$ is said to be matrix (or, operator) monotone (increasing) on an interval
$I\subset D_{f}\subset%
\mathbb{R}
$, with $D_{f}$ denoting the domain of definition of $f$, if for all Hermitian
matrices $A$ and $B$ of all orders whose eigenvalues lie in $I$, $A\geq
B\Rightarrow f\left(  A\right)  \geq f\left(  B\right)  $. Observe that $A\geq
B$ if and only if $A-B$ is a positive matrix. We point of that the concept of
an operator monotone function can be subtle. For instance, there are examples
of monotone functions that are not operator monotone (for instance
\cite{karol06}, $f\left(  t\right)  =t^{2}$). For more details on the notion
of operator monotone functions along with suitable techniques to construct
them, we refer to Refs.
\cite{lowner34,bhatia97,kwong89,furuta08,gibilisco09,simon19}. The key
contribution by Petz in Ref. \cite{petz96a} was that of using operator
monotone functions to construct explicit examples of monotone metrics. The
joint work of Morozova-Chentsov-Petz (MCP) led to the much appreciated MCP
theorem \cite{morozova91,petz96a}. Roughly speaking, this theorem states that
every monotone metric on the space of density matrices can be recast in a
suitable form specified by a so-called Morozova-Chentsov (MC) function. A
scalar function $f:%
\mathbb{R}
_{+}\rightarrow%
\mathbb{R}
_{+}$ is called Morozova-Chentsov if it satisfies three conditions: (i) $f$ is
operator monotone; (ii) $f$ is self inversive, that is $f\left(  1/t\right)
=f\left(  t\right)  /t$ for any $t\in%
\mathbb{R}
_{+}$; and (iii) $f\left(  1\right)  =1$. Condition (ii) is necessary to have
a symmetric mean $A\#B$ between two Hermitian operators $A$ and $B$
\cite{karol06}. Recall that $A\#B\overset{\text{def}}{=}\sqrt{A}f\left(
\frac{1}{\sqrt{A}}B\frac{1}{\sqrt{A}}\right)  \sqrt{A}$, where $A>0$ and $f$
is an operator monotone function on $\left[  0\text{, }\infty\right)  $ with
$f(1)=1$. Finally, condition (iii) is a normalization condition which helps to
avoid a conical singularity of the metric at the maximally mixed state.

In the next subsection, we do not discuss the non-monotonicity of the
Sj\"{o}qvist metric by providing the existence of a CPTP map that violates the
inequality in Eq. (\ref{jovi}). Rather, we argue that the Sj\"{o}qvist metric
is not a monotone metric because it violates the MCP theorem since it is not
specified by a proper Morozova-Chentsov function.

\subsection{Discussion}

Consider two neighboring single-qubit density matrices $\rho$ and $\rho+d\rho$
in the Bloch ball, with $\rho$ given by
\begin{equation}
\rho=\frac{\hat{1}+\vec{r}\cdot\vec{\sigma}}{2}=\frac{1}{2}\left(
\begin{array}
[c]{cc}%
1+r\cos\left(  \theta\right)  & r\sin\left(  \theta\right)  e^{-i\varphi}\\
r\sin\left(  \theta\right)  e^{i\varphi} & 1-r\cos\left(  \theta\right)
\end{array}
\right)  \text{,} \label{g1}%
\end{equation}
and a diagonal form specified by $\rho_{\mathrm{diag}}=\left(  1/2\right)
\mathrm{diag}\left(  1+r\text{, }1-r\right)  $. In Eq. (\ref{g1}), $\vec{r}$
denotes the polarization vector given by $\vec{r}\overset{\text{def}}{=}%
r\hat{n}$ with $\hat{n}\overset{\text{def}}{=}\left(  \sin\left(
\theta\right)  \cos\left(  \varphi\right)  \text{, }\sin\left(  \theta\right)
\sin\left(  \varphi\right)  \text{, }\cos\left(  \theta\right)  \right)  $.
Observe that for mixed quantum states, $0\leq r<1$ and $\det\left(
\rho\right)  =\left(  1/2\right)  \left(  1-\vec{r}^{2}\right)  \geq0$ because
of the positiveness of $\rho$. For pure quantum states, instead, we have $r=1$
and $\det\left(  \rho\right)  =0$. According to the MCP theorem, any
Riemannian monotone metric between $\rho$ and $\rho+d\rho$ in the Bloch ball
with $\rho$ in Eq. (\ref{g1}) can be recast as \cite{karol06}%
\begin{equation}
ds^{2}=\frac{1}{4}\left[  \frac{dr^{2}}{1-r^{2}}+\frac{1}{f\left(  \frac
{1-r}{1+r}\right)  }\frac{r^{2}}{1+r}d\Omega^{2}\right]  \text{,} \label{MCP}%
\end{equation}
with $0<r<1$. In Eq. (\ref{MCP}), $d\Omega^{2}\overset{\text{def}}{=}%
d\theta^{2}+\sin^{2}\left(  \theta\right)  d\varphi^{2}$ specifies the metric
on the unit $2$-sphere while $f:%
\mathbb{R}
_{+}\rightarrow%
\mathbb{R}
_{+}$ is the so-called Morozova-Chentsov function $f=f\left(  t\right)  $.
Note that at the maximally mixed state where $r=0$, $t$ is defined as
$t\left(  r\right)  \overset{\text{def}}{=}\left(  1-r\right)  /(1+r)\in
\left[  0\text{, }1\right]  $ and becomes $t\left(  0\right)  =1$. Therefore,
the constraint (iii) (i.e., $f\left(  1\right)  =1\neq0$) is necessary to
bypass a conical singularity in the metric. In the case of the Bures metric,
\begin{equation}
ds_{\mathrm{Bures}}^{2}=\frac{1}{4}\left[  \frac{dr^{2}}{1-r^{2}}+r^{2}%
d\Omega^{2}\right]  \text{.} \label{BuresMCP}%
\end{equation}
From Eqs. (\ref{MCP}) and (\ref{BuresMCP}),%
\begin{equation}
\frac{1}{f_{\mathrm{Bures}}\left(  \frac{1-r}{1+r}\right)  }\frac{r^{2}}%
{1+r}=r^{2}\text{.} \label{paul}%
\end{equation}
Then, recalling that\textbf{ }$r\left(  t\right)  \overset{\text{def}%
}{=}\left(  1-t\right)  /(1+t)$\textbf{, }we find from Eq. (\ref{paul}) that
that
\begin{equation}
f_{\mathrm{Bures}}\left(  t\right)  \overset{\text{def}}{=}\frac{1+t}%
{2}\text{.} \label{fbures}%
\end{equation}
Clearly, $f_{\mathrm{Bures}}\left(  t\right)  $ satisfies conditions (i),
(ii), and (iii) \cite{karol06}. In the case of the Sj\"{o}qvist metric, we
have%
\begin{equation}
ds_{\mathrm{Sj\ddot{o}qvist}}^{2}=\frac{1}{4}\left[  \frac{dr^{2}}{1-r^{2}%
}+d\Omega^{2}\right]  \text{.} \label{SjoqvistMCP}%
\end{equation}
From Eqs. (\ref{MCP}) and (\ref{SjoqvistMCP}), we find that \cite{cafaroprd22}%
\begin{equation}
f_{\mathrm{Sj\ddot{o}qvist}}\left(  t\right)  \overset{\text{def}}{=}\frac
{1}{2}\frac{\left(  1-t\right)  ^{2}}{1+t}\text{.} \label{ferik}%
\end{equation}
For a brief comparative discussion on Eqs. (\ref{BuresMCP}) and
(\ref{SjoqvistMCP}) along with remarks on finite lengths of geodesics
connecting mixed quantum states in the Bures and Sj\"{o}qvist geometries, we
refer to Appendix C. We observe now\textbf{ }that although $f_{\mathrm{Sj\ddot
{o}qvist}}\left(  t\right)  $ is self inversive since $f_{\mathrm{Sj\ddot
{o}qvist}}\left(  1/t\right)  =f_{\mathrm{Sj\ddot{o}qvist}}\left(  t\right)
/t$, $f_{\mathrm{Sj\ddot{o}qvist}}\left(  1\right)  =0\neq1$. Therefore, as
pointed out in Ref. \cite{erik20}, the Sj\"{o}qvist metric in Eq.
(\ref{SjoqvistMCP}) is singular at the origin of the Bloch ball where $r=0$
(i.e., $t\equiv t\left(  0\right)  =1$) and the angular components of the
metric tensor diverge because $f_{\mathrm{Sj\ddot{o}qvist}}\left(  1\right)
=0$. For this reason, the original Sj\"{o}qvist metric is limited to
non-degenerate mixed quantum states. Alternatively, the emergence of the
singular behavior of the Sj\"{o}qvist metric expressed in the form of Eq.
(\ref{MCP}) can be understood by noting that\textbf{ }$1/f_{\mathrm{Sj\ddot
{o}qvist}}\left(  \frac{1-r}{1+r}\right)  =(1+r)/r^{2}$\textbf{ }diverges
as\textbf{ }$r$\textbf{ }approaches zero\textbf{.} To properly understand the
monotonicity property of the Sj\"{o}qvist metric, we need to also check if
$f_{\mathrm{Sj\ddot{o}qvist}}\left(  t\right)  $ in Eq. (\ref{ferik}) is an
operator monotone function.

To address this point, we start by recalling that in spherical coordinates the
normalized volume element on the manifold of single-qubit mixed states
equipped with the most general Riemannian monotone metric is given by
\cite{karol06,petz1996}%
\begin{equation}
dV\overset{\text{def}}{=}p\left(  r\text{, }\theta\text{, }\varphi\right)
drd\theta d\varphi=\mathcal{N}\frac{r^{2}\sin\left(  \theta\right)  }{f\left(
\frac{1-r}{1+r}\right)  \left(  1-r^{2}\right)  ^{1/2}\left(  1+r\right)
}drd\theta d\varphi\text{,} \label{vol}%
\end{equation}
where $\mathcal{N}$ is a constant such that the probability density function
(pdf) $p\left(  r\text{, }\theta\text{, }\varphi\right)  $ in Eq. (\ref{vol})
is normalized to unity. For instance, in the Bures and Sj\"{o}qvist metric
cases, we have%
\begin{equation}
p_{\mathrm{Bures}}\left(  r\text{, }\theta\text{, }\varphi\right)
\overset{\text{def}}{=}\frac{1}{\pi^{2}}\frac{r^{2}\sin\left(  \theta\right)
}{\sqrt{1-r^{2}}}\text{, and }p_{\mathrm{Sj\ddot{o}qvist}}\left(  r\text{,
}\theta\text{, }\varphi\right)  \overset{\text{def}}{=}\frac{1}{2\pi^{2}}%
\frac{\sin\left(  \theta\right)  }{\sqrt{1-r^{2}}}\text{,} \label{vol1}%
\end{equation}
respectively. Note that from Eqs. (\ref{vol}) and (\ref{vol1}), $\mathcal{N}%
_{\mathrm{Bures}}\overset{\text{def}}{=}1/\pi^{2}$ and $\mathcal{N}%
_{\mathrm{Sj\ddot{o}qvist}}\overset{\text{def}}{=}1/(2\pi^{2})$. In Ref.
\cite{karol98}, Zyczkowski-Horodecki-Sanpera-Lewenstein (ZHSL) introduced a
\textquotedblleft natural measure\textquotedblright\ in the space of density
matrices specifying $N$-dimensional quantum systems to compute the volume of
separable and entangled states. The probability measure $\mu_{\mathrm{unitary}%
}$ used by ZHLS to describe the manner in which $N\times N$ random density
matrices $\rho$ that describe $N$-dimensional quantum systems are drawn, is
specified by means of a product $\mu_{\mathrm{unitary}}=\Delta_{1}\times
\nu_{\mathrm{Haar}}$. The quantity $\nu_{\mathrm{Haar}}$ denotes the Haar
measure in the space of unitary matrices $U(N)$
\cite{marinov80,maekawa85,boya03,alsing22}, while $\Delta_{1}$ is the uniform
measure on the $\left(  N-1\right)  $-dimensional simplex defined by the
constraint $\sum_{i=1}^{N}d_{i}=1$ (where $\left\{  d_{i}\right\}  _{1\leq
i\leq N}$ are the $N$ positive eigenvalues of $\rho$) \cite{chatterjee17}.
ZHLS proposed the product $\mu_{\mathrm{unitary}}=\Delta_{1}\times
\nu_{\mathrm{Haar}}$ motivated by the rotational invariance of both terms
$\Delta_{1}$ and $\nu_{\mathrm{Haar}}$. In Ref. \cite{karol99}, Zyczkowski
discussed the measure-dependence of questions concerning the separability of
randomly chosen mixed quantum states expressed as a mixture of pure states in
an $N$-dimensional Hilbert space. In Ref. \cite{slater00}, focusing on the
two-dimensional case with $N=2$, Slater showed that the pdf that characterizes
the ZHSL volume element equals%
\begin{equation}
p_{\mathrm{ZHSL}}\left(  r\text{, }\theta\text{, }\varphi\right)
\overset{\text{def}}{=}\frac{\Gamma\left(  \frac{1}{2}+\nu\right)  }%
{2\pi^{3/2}\Gamma\left(  \nu\right)  }\left(  1-r^{2}\right)  ^{\nu-1}%
\sin\left(  \theta\right)  \text{,}%
\end{equation}
where $\Gamma\left(  \nu\right)  $ is the Euler gamma function and $\nu>0$ is
the usual concentration parameter that appears in probability theory
\cite{sivia}. Recasting $dV_{\mathrm{ZHSL}}\overset{\text{def}}{=}%
p_{\mathrm{ZHSL}}\left(  r\text{, }\theta\text{, }\varphi\right)  drd\theta
d\varphi$ as in Eq. (\ref{vol}) and following Slater's work, we get%
\begin{equation}
f_{\mathrm{ZHSL}}\left(  t\text{; }\nu\right)  \overset{\text{def}%
}{=}\mathcal{N}_{\mathrm{ZHSL}}\left(  \nu\right)  \cdot\frac{2\pi^{3/2}%
\Gamma\left(  \nu\right)  }{\Gamma\left(  \frac{1}{2}+\nu\right)  }\cdot
\frac{1}{2}\frac{\left(  1-t\right)  ^{2}}{1+t}\cdot\left(  \frac{4t}{\left(
1+t\right)  ^{2}}\right)  ^{\frac{1}{2}-\nu}\text{.} \label{fkarol}%
\end{equation}
\bigskip In Ref. \cite{slater00}, Slater noticed that the one-parameter family
of functions $f_{\mathrm{ZHSL}}\left(  t\text{; }\nu\right)  $ in Eq.
(\ref{fkarol}) are such that $f_{\mathrm{ZHSL}}\left(  1\text{; }\nu\right)
=0\neq1$, for any $\nu>0$. Therefore, these functions are not normalizable as
required by a proper Morozova-Chentsov function. However, since
$f_{\mathrm{ZHSL}}\left(  1/t\text{; }\nu\right)  =f_{\mathrm{ZHSL}}\left(
t\text{; }\nu\right)  /t$, $f_{\mathrm{ZHSL}}\left(  t\text{; }\nu\right)  $
is self inversive. Furthermore, although $f_{\mathrm{ZHSL}}\left(  t\text{;
}\nu\right)  $ is monotone decreasing for $t\in\left[  0,1\right]  $ and
monotone increasing for $t>1$, they are not operator monotone \cite{slater00}.
Thus, $dV_{\mathrm{ZHSL}}$ is not proportional to the volume element of a
monotonic metric. As a consequence, any metric associated with the ZHSL
measure would lack the statistically meaningful feature of decreasing under
the action of stochastic mappings \cite{slater00,slater99}. Comparing Eqs.
(\ref{ferik}) and (\ref{fkarol}), for $\nu=1/2$ we have
\begin{equation}
f_{\mathrm{ZHSL}}\left(  t\text{; }1/2\right)  =f_{\mathrm{Sj\ddot{o}qvist}%
}\left(  t\right)  \text{,} \label{carluccio}%
\end{equation}
where $\mathcal{N}_{\mathrm{ZHSL}}\left(  1/2\right)  =1/(2\pi^{2}%
)=\mathcal{N}_{\mathrm{Sj\ddot{o}qvist}}$. Thus, exploiting the finding of
Slater in Refs. \cite{slater00,slater99}, we conclude that for $N=2$ the
Sj\"{o}qvist metric is not a monotone metric (unlike the Bures
metric).\textbf{ }For completeness, we point out that one can explicitly
verify that\textbf{ }$f_{\mathrm{Sj\ddot{o}qvist}}\left(  t\right)  $\textbf{
}in Eq. (\ref{ferik}) on\textbf{ }$\left[  0\text{, }1\right]  $ is not
operator monotone since there exist positive matrices\textbf{ }$A$\textbf{,
}$B$\textbf{ }such\textbf{ }that\textbf{ }$B-A$\textbf{ }is positive
but\textbf{ }$f_{\mathrm{Sj\ddot{o}qvist}}(B)-f_{\mathrm{Sj\ddot{o}qvist}}%
(A)$\textbf{ }is not. To see this, take\textbf{ }$B=I$\textbf{ }and\textbf{
}$A=I/2$\textbf{ }with\textbf{ }$I$\textbf{ }being the\textbf{ }$2\times
2$\textbf{ }identity matrix.\textbf{ }The discovery of the link in Eq.
(\ref{carluccio}) between the family of\textbf{ }$\mathrm{ZHSL}$\textbf{
}metrics and the Sj\"{o}qvist metric is intriguing in its own right and, we
believe, goes beyond the monotonicity aspects being discussed here. We are now
ready for our summary and\textbf{ }concluding remarks.

\section{Conclusion}

In this paper, we presented an explicit mathematical discussion on the link
between the Sj\"{o}qvist metric and the Bures metric for arbitrary
nondegenerate mixed quantum states in terms of decompositions of density
operators via ensembles of pure quantum states. Furthermore, to deepen our
physical understanding of the difference between these two metrics, we found
and compared the formal expressions of these two metrics for arbitrary thermal
quantum states describing quantum systems in equilibrium with an environment
at non-zero temperature (Eqs. (\ref{from3}) and (\ref{from4})). Finally, we
illustrated the discrepancy (Eq. (\ref{s3})) between these two metrics (Eqs.
(\ref{s1}) and (\ref{s2})) in the case of a simple physical system defined by
a spin-qubit in an arbitrarily oriented uniform and stationary magnetic field
in thermal equilibrium with a finite-temperature reservoir. Our main
conclusive remarks can be summarized as follows:

\begin{enumerate}
\item[{[i]}] Motivated by the original considerations presented in Ref.
\cite{erik20}, we have explicitly clarified that the Sj\"{o}qvist metric
$d_{\mathrm{Sj\ddot{o}qvist}}^{2}\left(  t\text{, }t+dt\right)  $ in Eq.
(\ref{cosmo4}) is generally different from the Bures metric $d_{\mathrm{Bures}%
}^{2}\left(  t\text{, }t+dt\right)  $ in Eq. (\ref{burino}).

\item[{[ii]}] Building on the quantitative analysis that appeared in Ref.
\cite{erik20}, we have explicitly verified that the generalized Sj\"{o}qvist
metric $\tilde{d}_{\mathrm{Sj\ddot{o}qvist}}^{2}\left(  t\text{, }t+dt\right)
$ in Eq. (\ref{cosmo17}) coincides with the Bures metric $d_{\mathrm{Bures}%
}^{2}\left(  t\text{, }t+dt\right)  $ in Eq. (\ref{burino}).

\item[{[iii]}] We have explicitly stated that $d_{\mathrm{Bures}}^{2}\left(
t\text{, }t+dt\right)  =\tilde{d}_{\mathrm{Sj\ddot{o}qvist}}^{2}\left(
t\text{, }t+dt\right)  \leq d_{\mathrm{Sj\ddot{o}qvist}}^{2}\left(  t\text{,
}t+dt\right)  $. This inequality is a consequence of the fact that in the
generalized Sj\"{o}qvist metric construction, the minimization procedure
occurs in a larger space of unitary matrices (Eq. (\ref{cos12})) that includes
the smaller space of unitary matrices (Eq. (\ref{cosmo3})) explored in the
original Sj\"{o}qvist construction.

\item[{[iv]}] Inspired by the work in Ref. \cite{erik20}\textbf{, }we have
explicitly point out that either $d_{\mathrm{Sj\ddot{o}qvist}}^{2}\left(
t\text{, }t+dt\right)  $ or $\tilde{d}_{\mathrm{Sj\ddot{o}qvist}}^{2}\left(
t\text{, }t+dt\right)  $ can be obtained starting from a common general
minimization procedure. However, these two metrics are generally different
since they correspond to different minima (i.e., different choices of the
unitary matrix $V\left(  t\right)  \leftrightarrow\left[  V_{hk}\left(
t\right)  \right]  _{1\leq h,k\leq N}$ with $V_{hk}\left(  t\right)  \in%
\mathbb{C}
$ introduced in Eq. (\ref{cosmo10})).

\item[{[v]}] For the class of thermal states in an arbitrary
finite-dimensional setting, we stressed the difference between the
Sj\"{o}qvist and the Bures metrics in terms of their non-classical metric
components (Eqs. (\ref{from3}) and (\ref{from4})).

\item[{[vi]}] For single-qubit mixed states, we argued that unlike the Bures
metric (with the MC function in Eq. (\ref{fbures})), the Sj\"{o}qvist metric
(with the MC-like function in Eq. (\ref{ferik})) is not a monotone metric.
\end{enumerate}

For the set of pure states there is no room for ambiguity and the
(unitary-invariant) Fubini--Study metric leads to the only natural choice for
a measure that defines \textquotedblleft random states\textquotedblright. For
mixed-state density matrices, instead, the geometric structure of the state
space is more intricate \cite{karol06,brody19}. There is a variety of
different metrics that can be employed, each of them with a different physical
justification, advantages, and drawbacks that can depend on the specific
application one \ might examine. In particular, both basic geometric
quantities (i.e., path, path length, volume, and curvature) and more involved
geometric concepts built out of these basic entities (i.e., complexity) happen
to depend on the measure chosen on the space of mixed quantum states that
specify the physical system being studied \cite{karol99,cafaroprd22}. For
these reasons, our work carried out in this paper can be especially relevant
in providing a clearer comparative analysis between the (younger) Sj\"{o}qvist
interferometric geometry and the (older) Bures geometry for mixed quantum
states. Interestingly, the relevance of this type of comparative analysis was
recently remarked in Refs. \cite{mera22} and \cite{cafaroprd22} as well.

It would be interesting to investigate the monotonicity of the Sj\"{o}qvist
metric for $N>2$. In particular, keeping $N=2$, it would be intriguing to
identify an explicit counterexample of a CPTP map for single-qubits for which
the Sj\"{o}qvist distance does not decrease under its action (see, for
instance, Ref. \cite{ozawa00} for the existence of an explicit counterexample
exhibiting the non-monotonicity of the Hilbert-Schmidt distance). Finally,
thanks to Eq. (\ref{carluccio}), we found for $N=2$ and $\nu=1/2$ that the
metric associated with the ZHSL measure is equal to the Sj\"{o}qvist metric in
Eq. (\ref{SjoqvistMCP}). This connection deserves further investigation, we
believe. For the time being, we leave a deeper quantitative understanding of
these lines of investigation to forthcoming scientific efforts.

Despite its relative simplicity, we hope this work will inspire other
scientists to strengthen our mathematical and physical comprehension of this
intriguing link among geometry, statistical mechanics, and quantum physics.

\begin{acknowledgments}
P.M.A. acknowledges support from the Air Force Office of Scientific Research
(AFOSR). C.C. is grateful to the United States Air Force Research Laboratory
(AFRL) Summer Faculty Fellowship Program for providing support for this work.
Any opinions, findings and conclusions or recommendations expressed in this
material are those of the author(s) and do not necessarily reflect the views
of the Air Force Research Laboratory (AFRL). O.L. is grateful to the
Department of Physics of the Al-Farabi University for hospitality during the
period in which this manuscript was written. C.L. and S.M. acknowledge
financial support from \textquotedblleft PNRR MUR project
PE0000023-NQSTI\textquotedblright. Finally, the work of H.Q. was supported
partially by PAPIIT-DGAPA-UNAM, Grant No. 114520, and Conacyt-Mexico, Grant
No. A1-S-31269.
\end{acknowledgments}

\bigskip\pagebreak

\appendix

\section{Parallel transport condition for mixed quantum states}

In Appendix A\textbf{,} to better grasp the significance of the relation
$U_{t}\left(  dt\right)  V\left(  t+dt\right)  V^{\dagger}\left(  t\right)
=I$ in Eq. (\ref{condition2}), we recall the concept of parallel transport for
pure \cite{aharonov87} and mixed \cite{erik00,tong04} quantum states.

Remember that a unitarily evolving mixed quantum state $\rho\left(  t\right)
\overset{\text{def}}{=}U\left(  t\right)  \rho\left(  0\right)  U^{\dagger
}\left(  t\right)  $ is said to gain a geometric phase with respect to
$\rho\left(  0\right)  $ if $\arg\left\{  \mathrm{tr}\left[  \rho\left(
0\right)  U\left(  t\right)  \right]  \right\}  $ is nonzero \cite{erik00}.
Then, the parallel transport condition of $\rho\left(  t\right)  $ along an
arbitrary path is specified by the condition that the state $\rho\left(
t\right)  $ must be, at each temporal interval, in phase with the state
$\rho\left(  t+dt\right)  \overset{\text{def}}{=}U\left(  t+dt\right)
\rho\left(  0\right)  U^{\dagger}\left(  t+dt\right)  =U\left(  t+dt\right)
U^{\dagger}\left(  t\right)  \rho\left(  t\right)  U\left(  t\right)
U^{\dagger}\left(  t+dt\right)  $. Being in phase requires, in turn, that
$\arg\left\{  \mathrm{tr}\left[  \rho\left(  t\right)  U\left(  t+dt\right)
U^{\dagger}\left(  t\right)  \right]  \right\}  $ must vanish, that is,
$\mathrm{tr}\left[  \rho\left(  t\right)  U\left(  t+dt\right)  U^{\dagger
}\left(  t\right)  \right]  $ must be \emph{real} and \emph{positive}.
However, noting that $U(d+dt)=U(t)+\dot{U}\left(  t\right)  dt+O\left(
dt^{2}\right)  $, the parallel transport condition can be recast as
$\arg\left\{  \mathrm{tr}\left[  \rho\left(  t\right)  \dot{U}\left(
t\right)  U^{\dagger}\left(  t\right)  \right]  \right\}  =0$. Finally, since
$\rho\left(  t\right)  \dot{U}\left(  t\right)  U^{\dagger}\left(  t\right)  $
is a purely imaginary number since $\rho=\rho^{\dagger}$ (Hermiticity) and
$UU^{\dagger}=U^{\dagger}U=\mathrm{I}$ (unitarity), the parallel transport
condition reduces to $\mathrm{tr}\left[  \rho\left(  t\right)  \dot{U}\left(
t\right)  U^{\dagger}\left(  t\right)  \right]  =0$. For a characterization of
the mixed state geometric phase in the case of nonunitary evolutions, we refer
to Ref. \cite{tong04}. For a pure state density operator $\rho\left(
t\right)  \overset{\text{def}}{=}\left\vert \psi\left(  t\right)
\right\rangle \left\langle \psi\left(  t\right)  \right\vert $, the parallel
transport condition is given by $\left\langle \psi\left(  t\right)  |\dot
{\psi}\left(  t\right)  \right\rangle =0$ \cite{aharonov87}. Therefore,
setting for example $\left\vert \psi\left(  t\right)  \right\rangle
\overset{\text{def}}{=}e^{if_{k}\left(  t\right)  }\left\vert n_{k}\left(
t\right)  \right\rangle $, the condition $\left\langle \psi\left(  t\right)
|\dot{\psi}\left(  t\right)  \right\rangle =0$ yields the scalar constraint
$\dot{f}_{k}\left(  t\right)  -i\left\langle n_{k}\left(  t\right)  |\dot
{n}_{k}\left(  t\right)  \right\rangle =0$ which was obtained by Sj\"{o}qvist
in his original derivation of the metric tensor for mixed quantum states.
Given this background information, we point out that the relation
$U_{t}\left(  dt\right)  V\left(  t+dt\right)  V^{\dagger}\left(  t\right)
=\mathrm{I}$ in Eq. (\ref{condition2}) is a constraint equation that can be
regarded as the operator-analogue of the parallel transport condition $\dot
{f}_{k}\left(  t\right)  -i\left\langle n_{k}\left(  t\right)  |\dot{n}%
_{k}\left(  t\right)  \right\rangle =0$. In particular, it is straightforward
to check that when the polar decomposition of the overlap matrix $M_{t}\left(
dt\right)  $ is given by $\left\vert M_{t}\left(  dt\right)  \right\vert
U_{t}\left(  dt\right)  $ with matrix coefficients $\left[  M_{t}\left(
dt\right)  \right]  _{kl}\overset{\text{def}}{=}\sqrt{p_{k}\left(  t\right)
p_{l}\left(  t\right)  }\left\langle n_{k}\left(  t\right)  |n_{l}\left(
t+dt\right)  \right\rangle $ that are diagonalizable with \emph{real} and
\emph{positive} eigenvalues, the relation $U_{t}\left(  dt\right)  V\left(
t+dt\right)  V^{\dagger}\left(  t\right)  =\mathrm{I}$ leads to the constraint
\textrm{tr}$\left[  \rho\left(  t\right)  \dot{V}\left(  t\right)  V^{\dagger
}\left(  t\right)  \right]  =0$. This latter relation can be explicitly
verified by exploiting the fact that \textrm{tr}$\left[  \rho\left(  t\right)
\right]  =1$ and, in addition, the unitary matrix $V\left(  t\right)  $
satisfies the relation $V(t+dt)=V(t)+\dot{V}\left(  t\right)  dt+O\left(
dt^{2}\right)  $. For a rigorous mathematical discussion on the notion of
parallel transport along density operators, we suggest Refs.
\cite{uhlman76,uhlmann86,uhlmann91,uhlmann92,uhlmann95}.

\section{Monotonicity of the Bures metric}

In this appendix, we report some details on the monotonicity property
satisfied by the Bures metric viewed as a Riemannian metric.

We recall that there exist infinitely many monotone Riemannian metrics on the
space of mixed quantum states \cite{karol06}. In particular, the monotonicity
of the Bures metric $d_{\mathrm{Bures}}^{2}\left(  t\text{, }t+dt\right)  $
under stochastic quantum maps $\left\{  \Phi\right\}  $ (i.e., completely
positive trace preserving (CPTP) maps) is a consequence of the monotonicity of
the Bures distance $D_{\mathrm{Bures}}^{2}\left(  \rho_{1}\text{, }\rho
_{2}\right)  $ \cite{karol06},%
\begin{equation}
D_{\mathrm{Bures}}^{2}\left(  \rho_{1}\text{, }\rho_{2}\right)
\overset{\text{def}}{=}\mathrm{tr}\left(  \rho_{1}\right)  +\mathrm{tr}\left(
\rho_{2}\right)  -2\mathrm{tr}(\sqrt{\rho_{2}^{1/2}\rho_{1}\rho_{2}^{1/2}%
})\text{,} \label{BD}%
\end{equation}
as a function of the fidelity $\mathrm{tr}(\sqrt{\rho_{2}^{1/2}\rho_{1}%
\rho_{2}^{1/2}})$. No physical operation expressed in terms of a CPTP map
$\Phi$ can increase $D_{\mathrm{Bures}}^{2}\left(  \rho_{1}\text{, }\rho
_{2}\right)  $,%
\begin{equation}
D_{\mathrm{Bures}}^{2}\left(  \Phi\rho_{1}\text{, }\Phi\rho_{2}\right)  \leq
D_{\mathrm{Bures}}^{2}\left(  \rho_{1}\text{, }\rho_{2}\right)  \text{.}
\label{con1}%
\end{equation}
To avoid confusion, we point out that the quantity $\mathrm{tr}(\sqrt{\rho
_{2}^{1/2}\rho_{1}\rho_{2}^{1/2}})$ is denoted with $\sqrt{F}\left(  \rho
_{1}\text{, }\rho_{2}\right)  $ and called root fidelity in Ref.
\cite{karol06} (see Eq. (9.33). Instead, $\mathrm{tr}(\sqrt{\rho_{2}^{1/2}%
\rho_{1}\rho_{2}^{1/2}})$ is denoted with $F\left(  \rho_{1}\text{, }\rho
_{2}\right)  $ and called fidelity in Ref. \cite{nielsen00} (see Eq. (9.53).
Interestingly, the fidelity $F\left(  \rho_{1}\text{, }\rho_{2}\right)  $ can
be used to define the so-called Bures angle $D_{A}^{\mathrm{Bures}}\left(
\rho_{1}\text{, }\rho_{2}\right)  $ as%
\begin{equation}
D_{A}^{\mathrm{Bures}}\left(  \rho_{1}\text{, }\rho_{2}\right)
\overset{\text{def}}{=}\arccos\left[  \mathrm{tr}(\sqrt{\rho_{2}^{1/2}\rho
_{1}\rho_{2}^{1/2}})\right]  \text{.} \label{BA}%
\end{equation}
The Bures angle $D_{A}^{\mathrm{Bures}}\left(  \rho_{1}\text{, }\rho
_{2}\right)  $ in Eq. (\ref{BA}) is a metric \cite{nielsen00} that, similarly
to the Bures distance $D_{\mathrm{Bures}}^{2}\left(  \rho_{1}\text{, }\rho
_{2}\right)  $ in Eq. (\ref{BD}), satisfies the contractivity property given
by%
\begin{equation}
D_{A}^{\mathrm{Bures}}\left(  \Phi\rho_{1}\text{, }\Phi\rho_{2}\right)  \leq
D_{A}^{\mathrm{Bures}}\left(  \rho_{1}\text{, }\rho_{2}\right)  \text{,}
\label{con2}%
\end{equation}
for any CPTP map $\Phi$ \cite{karol06}. Eq. (\ref{con2}) is a consequence of
two facts: i) $D_{A}^{\mathrm{Bures}}\left(  \rho_{1}\text{, }\rho_{2}\right)
$ in Eq. (\ref{BA}) is a monotone decreasing function of the fidelity
$\sqrt{\rho_{2}^{1/2}\rho_{1}\rho_{2}^{1/2}}$; ii) the fidelity $F\left(
\rho_{1}\text{, }\rho_{2}\right)  $, expressed as $\sqrt{\rho_{2}^{1/2}%
\rho_{1}\rho_{2}^{1/2}}$ and thanks to Uhlmann's theorem, can be shown to
fulfill the monotonicity property%
\begin{equation}
F\left(  \Phi\rho_{1}\text{, }\Phi\rho_{2}\right)  \geq F\left(  \rho
_{1}\text{, }\rho_{2}\right)  \text{,}%
\end{equation}
for any CPTP map $\Phi$ \cite{nielsen00}. For proof that fidelity does not
decrease under local general measurements (LGMs) and classical communication
(CC), we refer to Ref. \cite{barnum96}. Finally, for an interesting discussion
on the relevance of the contractivity property for distances used to properly
quantify entanglement in quantum information science, we refer to Refs.
\cite{vedral97,vedral98}.

\section{Finite lengths of geodesic paths}

We begin Appendix C by pointing out that in order to better understand from an
intuitive standpoint the difference between the Bures and the Sj\"{o}qvist
metrics, in addition to the expressions of their infinitesimal line elements
in Eqs. (\ref{BuresMCP}) and (\ref{SjoqvistMCP}), respectively, it would be
convenient to also have an explicit formula for the finite distance between
two arbitrary qubit mixed states. However, before addressing the problem of
finding the finite length of a geodesic path of suitably parametrized density
operators connecting an initial and a final mixed state, we present some
preliminary remarks. First, considering a change of variables defined by
$r\overset{\text{def}}{=}\sin\left(  \alpha_{r}\right)  $ with $0\leq
\alpha_{r}\leq\pi/2$, we obtain that $4ds_{\mathrm{Sj\ddot{o}qvist}}%
^{2}=d\alpha_{r}^{2}+d\Omega_{\text{sphere}}^{2}$ and $4ds_{\mathrm{Bures}%
}^{2}=d\alpha_{r}^{2}+\sin^{2}\left(  \alpha_{r}\right)  d\Omega
_{\text{sphere}}^{2}$ with $d\Omega_{\text{sphere}}^{2}\overset{\text{def}%
}{=}d\theta^{2}+\sin^{2}\left(  \theta\right)  d\varphi^{2}$. Recalling that
the line element in the standard cylindrical coordinates $\left(  \rho\text{,
}\varphi\text{, }z\right)  $ is given by $ds_{\text{cylinder}}^{2}%
=dz^{2}+d\Omega_{\text{cylinder}}^{2}$ with $d\Omega_{\text{cylinder}}%
^{2}\overset{\text{def}}{=}d\rho^{2}+\rho^{2}d\varphi^{2}$, one observes that
the structure of the Sj\"{o}qvist line element rewritten in this alternative
form is evocative of the structure of a line element in the standard\textbf{
}cylindrical coordinates once one associates the pair\textbf{ }$\left(
\alpha_{r}\text{, }d\Omega_{\text{sphere}}\right)  $\textbf{ }with the pair
$\left(  \rho\text{, }d\Omega_{\text{cylinder}}\right)  $. Second, after
considering this change of variables, one can connect a cylinder with a
constant (varying) radius to the Sj\"{o}qvist (Bures) geometry, respectively.
In particular, one observes that the varying radius in the Bures case is upper
bounded by the constant value that defines the radius in the Sj\"{o}qvist
geometry. These geometric insights would lead one to intuitively expect
different lengths of geodesic paths in the two cases, with the Sj\"{o}qvist
geometry yielding longer lengths eventually \cite{cafaroprd22}. Returning to
the issue of finite lengths, we consider for illustrative purposes two mixed
states $\rho_{A}$ and $\rho_{B}$ specified by Bloch vectors $\vec{a}=r_{a}%
\hat{n}_{a}$ and $\vec{b}=r_{b}\hat{n}_{b}$ with $\hat{n}_{a}%
\overset{\text{def}}{=}\left(  0\text{, }0\text{, }1\right)  $ and $\hat
{n}_{b}\overset{\text{def}}{=}\left(  \sin\left(  \theta_{b}\right)  \text{,
}0\text{, }\cos\left(  \theta_{b}\right)  \right)  $, respectively. In other
words, $\rho_{A}$ and $\rho_{B}$ are points in the Bloch sphere given in
spherical coordinates by $P_{A}=\left(  r_{a}\text{, }\theta_{a}\text{,
}\varphi_{a}\right)  \overset{\text{def}}{=}\left(  r_{a}\text{, }0\text{,
}0\right)  $ and $P_{B}=\left(  r_{b}\text{, }\theta_{b}\text{, }\varphi
_{b}\right)  \overset{\text{def}}{=}\left(  r_{b}\text{, }\theta_{b}\text{,
}0\right)  $,\textbf{ }respectively. Therefore, $\rho_{A}$ and $\rho_{B}$ are
assumed to be points that lie on the $xz$-plane since $\varphi_{a}=\varphi
_{b}=0$. A relatively straightforward calculation yields expressions of the
finite lengths evaluated along the geodesic paths connecting $\rho_{A}$ and
$\rho_{B}$ in the Bures and Sj\"{o}qvist cases, respectively. The lengths are
given by%
\begin{equation}
\mathcal{L}_{\mathrm{Bures}}\left(  r_{a}\text{, }r_{b}\text{, }\theta
_{b}\right)  =\left[  2\left\{  1-\sqrt{2\left[  \frac{1+r_{a}r_{b}\cos\left(
\theta_{b}\right)  }{4}+\sqrt{\frac{1-r_{a}^{2}}{4}\cdot\frac{1-r_{b}^{2}}{4}%
}\right]  }\right\}  \right]  ^{1/2}\text{,} \label{blength}%
\end{equation}
and \cite{erik20},%
\begin{equation}
\mathcal{L}_{\mathrm{Sj\ddot{o}qvist}}\left(  r_{a}\text{, }r_{b}\text{,
}\theta_{b}\right)  =\frac{1}{2}\sqrt{\theta_{b}^{2}+\left[  \arcsin\left(
r_{b}\right)  -\arcsin\left(  r_{a}\right)  \right]  }\text{,} \label{slength}%
\end{equation}
respectively. To further grasp insights into our discussion and, in addition,
to cross-check the consistency of our calculations with what is expected to
happen in the case of neighboring pure quantum states, we set\textbf{ }%
$r_{a}=r_{b}=1$\textbf{. }Then, Eqs. (\ref{blength}) and (\ref{slength})
reduce to%
\begin{equation}
\mathcal{L}_{\mathrm{Bures}}\left(  \theta_{b}\right)  =\left[  2\left(
1-\sqrt{\frac{1+\cos\left(  \theta_{b}\right)  }{2}}\right)  \right]
^{1/2}\text{, and }\mathcal{L}_{\mathrm{Sj\ddot{o}qvist}}\left(  \theta
_{b}\right)  =\frac{\theta_{b}}{2}\text{,} \label{cumpa}%
\end{equation}
respectively. From Eq. (\ref{cumpa}) we observe that $0\leq\mathcal{L}%
_{\mathrm{Bures}}\left(  \theta_{b}\right)  \leq\mathcal{L}_{\mathrm{Sj\ddot
{o}qvist}}\left(  \theta_{b}\right)  $ for any $0\leq\theta_{b}\leq\pi$.
Moreover, for neighboring quantum states with $\theta_{b}\ll1$,\textbf{ }the
second order Taylor expansions in $\theta_{b}$ of $\mathcal{L}_{\mathrm{Bures}%
}\left(  \theta_{b}\right)  $, $\mathcal{L}_{\mathrm{Sj\ddot{o}qvist}}\left(
\theta_{b}\right)  $, and $\mathcal{L}_{\mathrm{Fubini-Study}}\left(
\theta_{b}\right)  \overset{\text{def}}{=}(1/2)d_{\mathrm{Fubini-Study}%
}\left(  \theta_{b}\right)  $ with $d_{\mathrm{Fubini-Study}}\left(
\theta_{b}\right)  \overset{\text{def}}{=}2\left[  1-\cos^{2}(\theta
_{b}/2)\right]  ^{1/2}$ being the Fubini-Study distance \cite{cafarocqg23},
coincide. Indeed, when the Wootters angle $\theta_{b}\ll1$,\textbf{ }one finds
$\mathcal{L}_{\mathrm{Bures}}\left(  \theta_{b}\right)  \approx\mathcal{L}%
_{\mathrm{Sj\ddot{o}qvist}}\left(  \theta_{b}\right)  \approx\mathcal{L}%
_{\mathrm{Fubini-Study}}\left(  \theta_{b}\right)  \approx\theta_{b}/2$.

We defer a more in-depth quantitative comparison between the Sj\"{o}qvist and
Bures geometries based upon the difference between the finite lengths of
geodesics connecting arbitrary mixed quantum states to a future scientific endeavor.

\end{document}